\documentclass[11pt,a4paper]{article}

\usepackage{jheppub}
\usepackage[usenames,dvipsnames]{xcolor}
\usepackage{amssymb,bm,cancel} 
\usepackage{amsmath}
\usepackage{slashed}
\usepackage{mathtools}
\usepackage{amsfonts}    
\usepackage{dsfont}
\usepackage{pdfpages}
\usepackage{verbatim}
\usepackage{stmaryrd}
\usepackage{graphicx}
\usepackage{hyperref}
\usepackage{bbm}
\usepackage{tikz}
\usepackage{pgfplots}
\usetikzlibrary{intersections, pgfplots.fillbetween}
\usetikzlibrary{arrows}
\usetikzlibrary{decorations.pathmorphing}
\usetikzlibrary{ decorations.markings}
\usetikzlibrary{shapes.misc}
\tikzset{cross/.style={cross out, draw=black, minimum size=2*(#1-\pgflinewidth), inner sep=0pt, outer sep=0pt},
cross/.default={1pt}}
\usetikzlibrary{shapes.geometric}
\tikzset{
    partial ellipse/.style args={#1:#2:#3}{
        insert path={+ (#1:#3) arc (#1:#2:#3)}
    }
}

\renewcommand{\figurename}{Fig.}

\newcommand{\bi}{\begin{itemize}}
\newcommand{\ei}{\end{itemize}}
\newcommand{\bea}{\begin{eqnarray}}
\newcommand{\eea}{\end{eqnarray}}
\newcommand{\be}{\begin{equation}}
\newcommand{\ee}{\end{equation}}

\definecolor{pink}{rgb}{1.0, 0.13, 0.32}

\numberwithin{equation}{section}
\begin{document}

\vspace*{2.5cm}
\begin{center}
{ \LARGE \text{Sphere path integrals for fermionic gauge fields}\\}
\vspace*{1.7cm}
Chiara Baracco,$^1$ and Vasileios A. Letsios$^2$
\vspace*{0.6cm}
\vskip4mm
\begin{center}
{
\footnotesize
{
$^1$Department of Mathematics, King's College London, Strand, London WC2R 2LS, UK \\
$^2$Physique de l’Univers, Champs et Gravitation, UMONS,
Place du Parc 20, 7000 Mons, Belgium \\
}}
\end{center}

\begin{center}
{\textsf{\footnotesize{
chiara.baracco@kcl.ac.uk, vasileios.letsios@umons.ac.be}} } 
\end{center}

\vspace*{0.6cm}

\vspace*{0.6cm}


\end{center}
\vspace*{1.5cm}
\begin{abstract}
\noindent
We consider spin-$s \geq \frac{3}{2}$ complex, strictly massless fermionic gauge fields on the four-sphere, $S^4$, which is the Euclidean continuation of de Sitter spacetime, dS$_4$. For $s=\frac{5}{2}$, after briefly discussing different options for the quantisation of the theory on dS$_4$, we compute the one-loop sphere path integral. We explain how it can be expressed in terms of functional determinants, generalising previous results for $s=\frac{3}{2}$. We proceed to rewrite the result in terms of `bulk' unitary Harish-Chandra  characters in the discrete series of the de Sitter isometry group, $Spin(4,1)$, and `edge' characters. We then generalise the character expression of the sphere path integral for any complex, strictly massless spin-$s\geq\frac{3}{2}$ fermionic gauge field. Additionally, we compute the coefficient of the logarithmic divergence of the $S^4$ path integral for all spin-$s \geq \frac{3}{2}$, and we show that it is fully encoded by bulk and edge characters. We further show that at one loop no imaginary phase appears for strictly massless fermionic gauge fields, contrary to the case of bosons. Lastly, we confirm the exact one-loop cancellation between bulk and edge contributions for the case of an infinite tower of complex massless fermions of spins $s=\frac{1}{2}, \frac{3}{2},\dots$, observed recently.
\end{abstract}

\newpage

\tableofcontents

\section{Introduction}

The Euclidean gravitational path integral is a fundamental object in  quantum gravity, and its study was pioneered in the seminal work \cite{Gibbons:1976ue} by Gibbons and Hawking. In the case of black holes, the Euclidean path integral is known to encode information about the thermodynamic properties of the quantum gravitational field in the semi-classical loop expansion. It also captures the quantum entanglement entropy of the vacuum \cite{Ryu:2006bv,Lewkowycz:2013nqa}, as well as the emission of quantum information from black holes \cite{Almheiri:2019qdq,Penington:2019npb}. Moreover, it can be used to determine logarithmic corrections to the entropy of black holes, providing constraints on microscopic completions of the theory \cite{Sen:2012dw}.
\newline\newline
In the case of quantum gravity with a positive cosmological constant, the path integral of interest is the sphere partition function, since the round metric on the sphere $S^4$ is the dominant saddle of Euclidean Einstein gravity. It corresponds to the Euclidean continuation of de Sitter spacetime, dS$_4$. According to the Gibbons-Hawking proposal \cite{Gibbons:1976ue,Gibbons:1977mu}, the sphere path integral $\mathcal{Z}$ encodes the quantum-corrected de Sitter entropy $\mathcal{S}$, as
\begin{align}
    \mathcal{Z} = \exp{\mathcal{S}} ~.
\end{align}
The entropic interpretation of the sphere path integral is motivated by the following observation. Expanding $\mathcal{Z}$ around the sphere saddle as
\begin{align}
    \mathcal{Z}\approx \mathcal{Z}_{\text{tree}} \,\mathcal{Z}_{1-\text{loop}} ~,
\end{align}
one finds that the tree-level contribution gives the area law for the entropy associated with the cosmological horizon of the static patch \cite{Gibbons:1976ue}, namely
\begin{align}
    \mathcal{Z}_{\text{tree}} = e^{\frac{A}{4G}} ~,
\end{align}
where $A=4 \pi \ell^2$ is the area of the dS$_4$ horizon and $\ell$ is the dS$_4$ radius, and we are setting $\hbar=1=c$. Then, according to the Gibbons-Hawking proposal, the one-loop contribution $\mathcal{Z}_{1-\text{loop}}$, as well as higher loops, encode quantum corrections to the area law, and $\frac{A}{4G}$ can be interpreted as an entropy \cite{Gibbons:1977mu} in analogy with the black hole case. These higher loops can be systematically computed, see for instance \cite{Bandaru:2024qvv, Anninos:2020hfj, Anninos:2021ene, Smith:2026dae, Muhlmann:2021clm}.
\newline\newline
Our current understanding of the de Sitter entropy is not as complete as in the case of black holes. Nevertheless, the success of the Euclidean gravitational path integral as a tool in the case of black holes motivates us to study one-loop path integrals of quantum fields placed on an $S^4$. Leaving the theoretical question of the de Sitter entropy aside, one-loop sphere path integrals are interesting in and of themselves, as they are fundamental objects of study in QFT and perturbative quantum gravity in de Sitter. Indeed, they are diffeomorphism and locally field-redefinition invariant calculables. Such path integrals are also interesting from a mathematical point of view, due to the recently discovered connection with Harish-Chandra characters of the Lorentzian dS group and `edge' characters  \cite{Anninos:2020hfj}.
\newline\newline
Within the context of microscopic models for expanding spacetimes, a promising theory is higher-spin gravity on dS$_4$, which features strong evidence of being dual to a microscopic CFT-like theory in Lorentzian  signature \cite{Anninos:2011ui, Anninos:2012ft, Anninos:2017eib, De:2026shn, Lang:2026inn, David:2020ptn, Neiman:2017zdr, Hertog:2017ymy}. It is worth mentioning that both the `bulk' and the microscopic theories are independently defined, which is not the case for other dS/CFT-like dualities in four dimensions. From a Euclidean perspective, the sphere partition function of higher-spin gravity should match the partition function of a  suitable microscopic  theory. This was recently addressed in \cite{anninos2026ds4metamorphosis}.  In the case of minimal higher-spin gravity, which contains only even-spin bosons, the one-loop analysis of \cite{anninos2026ds4metamorphosis} revealed that the microscopic theory is captured by two copies of the $Sp(N)$ model on $S^3$. The two copies are `glued' together through path integration over conformal higher-spin gauge fields in three dimensions. The authors also studied the supersymmetric case of higher-spin gravity (see \cite{Sezgin:2012ag}), whereby the relevant microscopic gluing formula involves two copies of the $U(N)$ model with reversed statistics on $S^3$. Strikingly, they showed that the finite part of the one-loop corrected $S^4$ partition function is equal to $2^N$ divided by the volume of the higher-spin (super)group. Moreover, if one considers a $U(N)$ and a  $U(M)$ model, then the ratio of the corresponding partition functions is independent of the group volume, and the result is not only unambiguous but also provides evidence for a counting interpretation of the sphere path integral of supersymmetric higher-spin gravity. Relatedly, supersymmetric methods were a key factor in formulating the counting interpretation of the Bekenstein-Hawking entropy \cite{Strominger:1996sh}. For recent works exploring the incorporation of supersymmetry in dS see  \cite{Boulanger:2026wnw, Higuchi-Letsios, Anninos:2023exn, Sezgin:2012ag, Pilch:1984aw, Chen:2025foq, Letsios:2023tuc, Anous:2014lia, Deser:2003gw, Hertog:2017ymy, Lang:2026inn}.  For further investigations concerning the sphere path integral, its  entropic interpretation and the interpretation of the `edge modes', see \cite{Maldacena:2024spf, Ivo:2025yek, Law:2025ktz, Chen:2025jqm, Giombi:2026sqa, Letsios:2026ypo, RiosFukelman:2023mgq, Law:2020cpj, Bobev:2022lcc, Jacobson:2022jir, Anninos:2025mje}.
\newline\newline
Our present paper is concerned with one-loop path integrals of massless higher-spin fermions on $S^4$. Our results are relevant to the proposal of \cite{anninos2026ds4metamorphosis}, since supersymmetric higher-spin gravity contains a tower of massless complex bosons (with spins $s=0,1,2, \dots $), as well as their massless fermionic superpartners ($s=\tfrac{1}{2} , \tfrac{3}{2}, \dots $).
\newline
\paragraph{Outline and results.} In section \ref{lor vs eucl section}, we begin by reviewing the theory of spin-$\frac{5}{2}$ strictly massless gauge fields on both Lorentzian and Euclidean backgrounds. We then proceed with the core sections of the paper, and each of them focuses on a clear, distinct target. First, in section \ref{spin-5/2 section} we compute the one-loop path integral for the complex, strictly massless spin-$\tfrac{5}{2}$ gauge field on $S^4$. We give a detailed analysis on how to express the path integral as a product of functional determinants and we take special care of the zero modes and the volume of residual gauge transformations. Secondly, in section \ref{spin-s section}, using our functional determinant form for the spin-$\frac{5}{2}$ sphere path integral, we rewrite the latter in terms of bulk unitary Harish-Chandra characters, $\chi_{\text{bulk}}^{(s)}$, of $Spin(4,1)$, i.e. the double cover of $SO(4,1)$, as well as `edge'  characters, $\chi_{\text{edge}}^{(s)}$. Moreover, we generalise the aforementioned character formula to the case of complex, strictly massless fermions of any spin-$s \geq \frac{5}{2}$ on $S^4$ (the spin-$\frac{3}{2}$ field has been already analysed in \cite{Anninos:2025mje}). We find that the one-loop $S^4$ path integral $\mathcal{Z}$ of any complex, strictly massless spin-$s \geq \tfrac{3}{2}$  fermion can be expressed as
\begin{equation}
    \log \mathcal{Z} = \log \left( \frac{\mathcal{Z}_{F^0,s}}{{\text{vol} \, \mathcal{G}_{\text{res}}}} \right) - \int_{\mathbb{R}^+} \frac{dt}{t} \frac{e^{-\frac{t}{2}}}{1-e^{-t}} ~2 \left( \chi_{\text{bulk}}^{(s)}(t) - \chi_{\text{edge}}^{(s)}(t) \right) ~,
\end{equation}
and further details can be found in section \ref{spin-s section}. This formula generalises the one of \cite{Anninos:2020hfj}, which concerns the case of strictly massless bosons. We also show that the unambiguous logarithmic divergence of the sphere path integral is purely encoded by the bulk and edge characters.
\newline\newline
For all strictly massless spin-$s \geq \tfrac{3}{2}$ fields, we show that no complex phase appears in the one-loop sphere path integral, which means that there is no fermionic analogue of Polchinski's phase for these theories, contrary to the case of strictly massless spin-$s \geq 2$ bosons \cite{Polchinski:1988ua,Anninos:2020hfj,Law:2020cpj}. This happens because there is no need to rotate integration contours for fermionic path integrals (contrary to the bosonic case), and a complex phase can only arise from the careful treatment of zero modes of the Euclidean action, since each zero mode gives a contribution proportional to $i$. As we later explain, the degeneracy of such zero modes is always an integer multiple of 4, which means that no complex phase appears since $i^4=1$. Lastly, we show that the sphere path integral for an infinite tower of massless higher-spin fields ($s=0, \tfrac{1}{2}, \tfrac{3}{2}, \dots $) on $S^4$ showcases exact cancellations between bulk and edge contributions, verifying earlier findings \cite{anninos2026ds4metamorphosis}.


\section{Lorentzian vs Euclidean theory} \label{lor vs eucl section}

In this brief section, we review the basics for the theory of a symmetric rank-2 tensor-spinor describing a strictly massless spin-$\frac{5}{2}$ gauge field on both dS$_4$ and its Euclidean continuation, $S^4$. As we will shortly explain, in Lorentzian signature the quantisation of the theory is not yet fully settled. In particular, although the Lorentzian quantisation has not been studied in detail, it is reasonable to expect that the theory will face issues akin to those of the massless spin-$\tfrac{3}{2}$ field \cite{Anninos:2025mje, Higuchi-Letsios, Letsios:2022tsq}. Nevertheless, at the level of the sphere path integral, the Euclidean theory seems to be straightforward to study, and the path integral itself is expressed in terms of unitary Harish-Chandra characters of $Spin(4,1)$, similarly to the spin-$\frac{3}{2}$ case \cite{Anninos:2025mje}. This result is interesting and somewhat surprising, given that in Lorentzian signature the canonical quantisation of the theory, carried out employing a Hermitian action, is expected to give rise to opposite norms for states of opposite helicity \cite{Letsios:2022tsq, Higuchi-Letsios}. The sphere path integral of the spin-$\frac{5}{2}$ theory will be examined in the rest of the paper.

\subsection{Lorentzian formulation}

Consider a spin-$\tfrac{5}{2}$ field described by a symmetric rank-2 tensor-spinor $\Psi_{\mu \nu} = \Psi_{ \nu \mu}$, with spinor indices suppressed, on a four-dimensional de Sitter (dS$_4$) background. The cosmological constant is $\Lambda = 3 {\ell}^{-2}$, where $\ell$ is the dS$_4$ radius. The equation of motion for the strictly massless (i.e. gauge potential) spin-$\tfrac{5}{2}$ field is
\begin{align} \label{eq 5/2}
    \mathcal{R}_{\mu \nu} [\Psi] =0 ~,
\end{align}
where \cite{Deser_2001, Fang:1979hq}
\begin{multline} \label{eq 5/2}
    \mathcal{R}_{\mu \nu} [\Psi]  = \slashed{\nabla} \Psi_{\mu\nu} + g_{\mu\nu} \Bigl( \gamma^\rho \nabla^\sigma \Psi_{\rho\sigma} - \frac{1}{2} \slashed{\nabla} \Psi_\rho^{~\rho} \Bigr) + \gamma_{(\mu} \bigl( \nabla_{\nu)} \Psi_\rho^{~\rho} + 2\slashed{\nabla} \gamma^\rho \psi_{\nu)\rho} \\
    - 2\nabla^\rho \Psi_{\nu)\rho} \bigr) - 2\nabla_{(\mu} \gamma^\rho \Psi_{\nu)\rho} + \frac{2i}{\ell} \Bigl( \Psi_{\mu\nu} - 2\gamma_{(\mu} \gamma^\rho \Psi_{\nu)\rho} - \frac{1}{2} g_{\mu\nu} \Psi_\rho^{~\rho} \Bigr) ~.
\end{multline}
Note that the field equation does not admit Majorana solutions.%
\footnote{Let us introduce the charge conjugate of $\Psi_{\mu \nu}$, which is defined as \cite{Freedman:2012zz}
\begin{align} \label{def: charge conjugate Lorentzian}
    \Psi_{\mu \nu}^C = B^{-1} \left( \Psi^{\dagger}_{\mu \nu} \right)^{T} ~,
\end{align}
where $B$ is a $4 \times 4$ spinorial matrix that satisfies
\begin{align}
    - (\gamma^{\mu~\dagger})^{T} = B \gamma^\mu B^{-1} ~.
\end{align}
It is easy to  check that if $\Psi_{\mu \nu}$ satisfies the field equation $\mathcal{R}_{\mu \nu} [\Psi] =0$, then so does $\Psi_{\mu \nu}^C$.  Majorana solutions of the field equation are those $\Psi_{\mu \nu} $'s that satisfy $\Psi^{C}_{\mu \nu} = \Psi_{\mu \nu}$. However, due the properties of the matrix $B$ \cite{Freedman:2012zz}, one finds $(\Psi_{\mu \nu}^C)^C=-\Psi_{\mu \nu}$. This means that all such Majorana tensor-spinors vanish identically. In other words, we cannot use the charge conjugation (\ref{def: charge conjugate Lorentzian}) in order to define a consistent reality (Majorana) condition. However, note that if one doubles the number of tensor-spinors, $\Psi^{(a)}_{\mu \nu}$, with $a=1,2$, then a consistent symplectic Majorana condition can be introduced, $\Psi^{(a)\,C}_{\mu \nu} = \epsilon^{ab} B^{-1} \left(\Psi^{(b)\,\dagger}_{\mu \nu} \right)^{T}$, where $\epsilon^{ab} = - \epsilon^{ba}$ (see, e.g., \cite{Freedman:2012zz, Pilch:1984aw}).  
See also \cite{Higuchi-Letsios} for a discussion concerning the lack of Majorana solutions for the strictly massless spin-$\tfrac{3}{2}$ field on dS$_4$. }
Here, the $\gamma^\mu$'s are the four gamma matrices satisfying the Clifford algebra anti-commutation relations in 4 dimensions, and $g_{\mu \nu}$ is the metric tensor of dS$_4$. The Dirac operator is $\slashed{\nabla} = \gamma^{\alpha}\nabla_{\alpha}$, where $\nabla_{\alpha}$ is the covariant derivative acting on tensor-spinors  --  see Appendix \ref{appendix formulae} for our spinor conventions.  Note that the mass parameter in (\ref{eq 5/2}), $\frac{2i}{\ell} $, is imaginary, analogously to the case of the strictly massless spin-$\tfrac{3}{2}$ field (gravitino) \cite{Letsios:2022tsq, Letsios:2023awz, Letsios:2023qzq, Higuchi-Letsios, Anninos:2025mje, Deser:2003gw, Cruz:2026odf}. It is easy to show that $\mathcal{R}_{\mu \nu} [\Psi]$ is invariant under the gauge transformation
\begin{equation} \label{gauge inv 5/2 lorentzian}
    \delta \Psi_{\mu\nu} = \nabla_{(\mu} \epsilon_{\tilde{\nu})} + \frac{i}{2\ell}\gamma_{(\mu} \epsilon_{\tilde{\nu})} ~, 
\end{equation}
where $\epsilon_{\nu}$ is a vector-spinor gauge parameter. The tilde above the vector index $\nu$  in $\epsilon_{\tilde{\nu}}$ denotes the $\gamma$-traceless part of $\epsilon_{{\nu}}$, i.e. $\! \epsilon_{\tilde{\nu}} \equiv \epsilon_\nu - \frac{1}{4} \gamma_\nu \gamma^\rho\epsilon_\rho$ and $\gamma^\nu \epsilon_{\tilde{\nu}} = 0$. 
\newline\newline
We now wish to address the question of what the action functional for this theory is. In the case of AdS$_4$ the answer is known, see \cite{Fang:1979hq,Campoleoni_2017}. By flipping the sign of the cosmological constant in the AdS$_4$ action for a Dirac spin-$\frac{5}{2}$ field \cite{Fang:1979hq}, so as to analytically continue the theory to dS$_4$, one finds
\begin{align} \label{spin 5/2 action naive}
    S_{\text{naive}} = -\int d^4x \sqrt{-g} ~\Psi^{\mu\nu\,\dagger}i \gamma^0~ \mathcal{R}_{\mu \nu} [\Psi] ~,
\end{align}
where $\Psi^{\mu\nu\,\dagger}i \gamma^0$ is the Dirac conjugate of $\Psi_{\mu \nu}$. Contrary to the AdS$_4$ case, the action (\ref{spin 5/2 action naive}) is complex due to the imaginary mass parameter of the spin-$\tfrac{5}{2}$ field in dS$_4$. Moreover, the above action is not invariant under the gauge variation of $\Psi^{\mu \nu \, \dagger}$. This problem is not specific to the spin-$\tfrac{5}{2}$ field, it also occurs if one naively continues the AdS$_4$ gravitino action to dS$_4$, see \cite{Higuchi-Letsios}. As for the gravitino \cite{Higuchi-Letsios, Boulanger:2026wnw}, a real, gauge invariant action does exist, and it can be obtained from (\ref{spin 5/2 action naive}) by inserting an extra factor of $\gamma^5$:
\begin{align}\label{spin 5/2 action gamma^5}
    S_{\gamma^5} = -\int d^4x \sqrt{-g} ~\Psi^{\mu\nu\,\dagger} i \gamma^0 \gamma^5 \, \mathcal{R}_{\mu \nu} [\Psi] ~.
\end{align}
Let us conclude by briefly discussing the quantisation of the field. One might think that the imaginary mass parameter indicates that the theory is ill-defined. Interestingly, it was recently shown in \cite{Letsios:2022tsq, Letsios:2022tsq, Letsios:2023awz} that the physical mode solutions of the field equations of  strictly massless tensor-spinor gauge potentials that can be obtained via analytic continuation from $S^4$, where (\ref{eq 5/2}) is a special case, furnish a direct sum of discrete series Unitary Irreducible Representations (UIRs) of the dS$_D$ isometry group $Spin(D,1)$, only for $D=4$.%
\footnote{There also exist fermionic discrete series UIRs of $Spin(2,1) \cong SL(2, \mathbb{R})$ that can be realised in the solution space of the Dirac equation for imaginary mass spinor fields on dS$_2$ \cite{Letsios-Fukelman}.  Interestingly, such discrete series spinors seem to be related to unconventional strongly coupled theories on dS$_2$ \cite{Galati:2026btt}. For a recent review on the representation theory of $Spin(D,1)$, $D \geq 2$, with emphasis on de Sitter field theory, see \cite{Hinterbichler:2026xqf}.} 
The existence of strictly and partially massless fermionic UIRs of $Spin(4,1)$ in the discrete series was already known in the mathematics literature \cite{ottoson1968classification, schwarz1971unitary}, but the fact that they can be realised in the solution space of imaginary mass tensor-spinor gauge fields on dS$_4$ was shown in  \cite{Letsios:2022tsq, Letsios:2022tsq, Letsios:2023awz}.
However, the existence of appropriate UIRs is a necessary but not sufficient condition for the formulation of a well-defined QFT.
Although the quantisation of the strictly massless spin-$\tfrac{5}{2}$ field theory on (Lorentzian) dS$_4$ is missing from the literature, we can draw  conclusions by comparing with the case of the strictly massless spin-$\tfrac{3}{2}$ QFT on dS$_4$ \cite{Higuchi-Letsios, Anninos:2025mje}. We distinguish (at least) three cases:
\begin{itemize}
    \item if we use the complex action (\ref{spin 5/2 action naive}), one can show that the resulting equations of motion are not self-consistent. The arguments are the same as the ones in the case of the gravitino, discussed in section 3 of \cite{Higuchi-Letsios};
    
    \item if we use the Hermitian action (\ref{spin 5/2 action gamma^5}), then we can write down a free QFT for the complex, strictly massless spin-$\tfrac{5}{2}$ field. This QFT includes the two discrete series UIRs in its single-particle Hilbert space, corresponding to the two helicities $\pm\tfrac{5}{2}$. However, the two fixed-helicity sectors have opposite norms because the $\gamma^5$ in (\ref{spin 5/2 action gamma^5}) is inherited from the natural dS invariant inner product \cite{Letsios:2022tsq, Letsios:2023awz}. Although the two distinct UIRs are realised separately in the fixed-helicity sectors with different definitions of positive-definite inner product, the resulting QFT is non-unitary because of the indefiniteness of the norm. Another drawback of this theory is that its flat space limit does not describe a standard Minkowskian massless spin-$\tfrac{5}{2}$ field, as the latter does not conventionally include a factor of $\gamma^5$ in its action.
    A possible way out in order to achieve a unitary QFT on dS$_4$ is to consider a chiral, strictly massless spin-$\tfrac{5}{2}$ theory, where the field strength is anti-self-dual, as explained in \cite{Higuchi-Letsios} for the gravitino. However, the anti-self-duality constraint is imposed on-shell and its incorporation at the level of the action is unclear; 

    \item if we drop the requirement that the field theory has an action functional, we can follow the arguments of \cite{Anninos:2025mje} for the gravitino and focus on quantisation using the field equations. In particular, we expect that one can identify appropriate field and conjugate momentum  operators on dS$_4$ such that the equal-time anti-commutators have the desired flat-space limit. Moreover, we expect positivity of the norm for both helicities because the associated inner product is different from the one stemming from the action (\ref{spin 5/2 action gamma^5}), as in the case of the gravitino \cite{Anninos:2025mje}.
\end{itemize}
The discussions above also apply to the quantisation of strictly massless fermions on dS$_4$ with any spin-$s \geq \tfrac{3}{2}$.


\subsection{Euclidean formulation}

We now place the theory on the four-sphere $S^4$, which is the Euclidean continuation of dS$_4$. Our aim is to compute the sphere path integral for the strictly massless spin-$\tfrac{5}{2}$ field. We choose the following Euclidean action
\begin{equation}\label{spin 5/2 action Euclidean complex}
 S_E[\Psi^{(E)}_{\mu\nu},\Xi^{\dagger}_{\mu\nu}] = \int_{S^4} d^4x \sqrt{g} ~\Xi^{\mu\nu \, \dagger} \mathcal{R}_{\mu \nu} [\Psi^{(E)}] ~.
\end{equation}
Here, $\Xi^{\mu \nu}$ and $\Psi^{(E)}_{\mu \nu}$ are two independent complex symmetric tensor-spinors, while $\Xi^{\mu \nu \, \dagger}$ denotes the Hermitian conjugate of  $\Xi^{\mu \nu}$, i.e. the tensor-spinor obtained from $\Xi^{\mu\nu}$ by taking the complex conjugate of all its components and simultaneously taking the transpose with respect to its spinor indices. With a slight abuse of notation, we will now drop the label $^{(E)}$ from $\Psi^{(E)}_{\mu \nu}$, and we will use $\Psi_{\mu \nu}$ to denote the Euclidean tensor-spinor. For our analysis, we choose the path integration contour whereby $\Xi_{\mu\nu} = \Psi_{\mu\nu}$. With this choice, the Euclidean action becomes
\begin{equation}\label{spin 5/2 action Euclidean}
    S_E[\Psi_{\mu\nu},\Psi^{\dagger}_{\mu\nu}] \equiv S_E[\Psi_{\mu\nu}] = \int_{S^4} d^4x \sqrt{g} ~\Psi^{\mu\nu \, \dagger} \mathcal{R}_{\mu \nu} [\Psi] ~,
\end{equation}
and it is invariant under the gauge transformation 
\begin{equation} \label{gauge inv 5/2}
    \delta \Psi_{\mu\nu} = \nabla_{(\mu} \epsilon_{\tilde{\nu})} + \frac{i}{2\ell}\gamma_{(\mu} \epsilon_{\tilde{\nu})} ~,
\end{equation}
as well as under the corresponding gauge transformation for $\Psi_{\mu \nu}^\dagger$. Contrary to the Lorentzian case, we do not need to add factors of $\gamma^5$ to ensure that the action is gauge invariant. In Euclidean signature we have $(\gamma^\mu)^\dagger = \gamma^\mu $ for all gamma matrices -- see Appendix \ref{appendix formulae} for more details. \\\\
Note that it is possible to derive the Euclidean action (\ref{spin 5/2 action Euclidean complex}) via analytic continuation from either of the two Lorentzian actions (\ref{spin 5/2 action naive}), (\ref{spin 5/2 action gamma^5}) discussed previously. When we make the transition from the Lorentzian to the Euclidean theory, the two Lorentzian tensor-spinors $\Psi^{\mu \nu \, \dagger}$ and $\Psi_{\mu \nu}$ in (\ref{spin 5/2 action naive}) and (\ref{spin 5/2 action gamma^5}) are respectively complexified as
\begin{equation}\label{analytic continuation rules naive}
    \Psi^{\mu \nu \, \dagger} i \gamma^0   \rightarrow  -\Xi^{\mu \nu \, \dagger} ~,~~~ \Psi_{\mu \nu} \rightarrow \Psi^{(E)}_{\mu \nu} ~,
\end{equation}
and
\begin{equation}\label{analytic continuation rules gamma5}
    \Psi^{\mu \nu \, \dagger} i \gamma^0 \gamma^5  \rightarrow  -\Xi^{\mu \nu \, \dagger} ~,~~~ \Psi_{\mu \nu} \rightarrow \Psi^{(E)}_{\mu \nu} ~,
\end{equation}
where we have reinserted the label $\,^{(E)}$ for clarity.


\section{Sphere path integral for spin-$\tfrac{5}{2}$ gauge field} \label{spin-5/2 section}

The aim of this section is to compute the one-loop sphere path integral for the strictly massless spin-$\tfrac{5}{2}$ gauge field on $S^4$,
\begin{align} \label{sphere path integral 5/2}
    \mathcal{Z} = \frac{1}{\text{vol} \, \mathcal{G}}\int D\Psi \,D\Psi^{\dagger}\, e^{-S
    _E[\Psi_{\mu\nu}]} ~,
\end{align}
where $\text{vol} \, \mathcal{G}$ denotes the volume of the gauge group associated with the gauge transformation (\ref{gauge inv 5/2}). In particular, we give a detailed analysis on how to express the path integral as a product of functional determinants. In section \ref{spin-s section}, we will rewrite the path integral in terms of bulk Harish-Chandra characters in the discrete series of $Spin(4,1)$, i.e. the double cover of $SO(4,1)$, as well as edge characters, and we will generalise the result to arbitrary spin-$s \geq 3/2$ strictly massless fermions. In what follows, we work on the unit $S^4$ by setting $\ell=1$.

\subsection{Field decomposition}

In order to compute the path integral (\ref{sphere path integral 5/2}), we begin by decomposing the spin-$\tfrac{5}{2}$ field into a transverse-traceless (TT) and a longitudinal part, as
\begin{align} \label{psi decomposition 2 generic}
    \Psi_{\mu\nu} & = \Psi_{\mu\nu}^{(\text{TT})} + \Psi_{\mu\nu}^{(\text{long})} ~,~~~ \text{with} ~~~ \nabla^{\mu}\Psi_{\mu\nu}^{(\text{TT})} = 0 = \gamma^{\mu}\Psi_{\mu\nu}^{(\text{TT})} ~.
\end{align}
Both the TT and the longitudinal components are complex (Dirac) tensor-spinors, and they are orthogonal to each other. The TT (/longitudinal) part can be expressed as a mode sum over TT (/longitudinal) spin-$\tfrac{5}{2}$ tensor-spinor spherical harmonics on $S^4$ \cite{Letsios:2022tsq}. We can further expand $\Psi_{\mu\nu}^{(\text{long})}$ and re-express the full field as
\begin{align} \label{psi decomposition 2}
    \Psi_{\mu\nu} 
    & = \Psi_{\mu\nu}^{(\text{TT})} + \nabla_{(\mu} \lambda_{\nu)} + \gamma_{(\mu} \chi_{\nu)} + g_{\mu\nu} B ~,~~~ \text{with} ~~~ \gamma^{\mu}\lambda_{\mu} = 0 = \gamma^{\mu}\chi_{\mu} ~,
\end{align}
where $\lambda_\mu$, $\chi_\mu$ are gamma-traceless Dirac vector-spinors and $B$ is a Dirac spinor. To ensure that the above expression is not redundant, let us further decompose the vector-spinors into their TT and longitudinal components (see \cite{Anninos:2025mje} for a detailed analysis of spin-$\tfrac{3}{2}$ gauge fields): 
\begin{equation}
    \lambda_\mu = \lambda_\mu^{(\text{TT})} + \left( \nabla_\mu - \frac{1}{4} \gamma_\mu \slashed{\nabla} \right) \xi ~,~~~ \text{with} ~~~ \nabla^{\mu}\lambda_{\mu}^{(\text{TT})} = 0 = \gamma^{\mu}\lambda_{\mu}^{(\text{TT})} ~,
\end{equation}
\begin{equation} \label{chi decomposition}
    {\chi}_\mu = {\chi}_\mu^{(\text{TT})} + \left( \nabla_\mu - \frac{1}{4} \gamma_\mu \slashed{\nabla} \right) \rho ~, ~~~ \text{with} ~~~ \nabla^{\mu}\chi_{\mu}^{(\text{TT})} = 0 = \gamma^{\mu}\chi_{\mu}^{(\text{TT})} ~.
\end{equation}
Then, to avoid redundancy in the decomposition (\ref{psi decomposition 2}) of $\Psi_{\mu \nu}$, we require that: 
\begin{itemize}
    \item $\lambda_\mu^{(\text{TT})}$ is orthogonal to Killing vector-spinors,\footnote{The logic behind the requirement is the following. We know that Killing vector-spinors are given by the spherical harmonics satisfying $\slashed{\nabla} \lambda_\mu^{(\text{TT})} = \pm 3i \lambda_\mu^{(\text{TT})}$, and equivalently they satisfy $\nabla_{(\mu} \lambda_{\nu)}^{(\text{TT})} = \pm \frac{i}{2} \gamma_{(\mu} \lambda_{\nu)}^{(\text{TT})}$. This means that, for Killing vector-spinor modes only, the terms involving $\lambda_\nu^{(\text{TT})}$ and $\chi_\nu^{(\text{TT})}$ in (\ref{psi decomposition 2}) appear with the same covariant structure, since $\nabla_\mu$ and $\gamma_\mu$ are essentially interchangeable. We then need only one between $\lambda_\nu^{(\text{TT})}$ and $\chi_\nu^{(\text{TT})}$ to carry such Killing vector-spinor modes, in order to reproduce the original field $\Psi_{\mu\nu}$. The same reasoning has been applied to determine which spinor modes must be excluded.} $\int_{S^4} d^4x\sqrt{g} \, \lambda^{(\text{TT}) \, \mu \, \dagger} A^{\pm}_{\mu} = 0$. Here, $A^{\pm}_{\mu}$ is a Killing vector-spinor, i.e. a solution of
    \begin{align} \label{def: Killing vec-spinors very nice}
        \nabla_{(\mu}A^{\mp}_{\nu)} \pm \frac{i}{2}\gamma_{(\mu} A^{\mp}_{\nu )} = 0 ~. 
    \end{align}
    For further details, see Appendix \ref{App: conformal Killing vec-spin};
    \item two out of the three spinors $\xi$, $\rho$, $B$ are  orthogonal to Killing spinors (we choose $\xi$ and $\rho$), $\int_{S^4} d^4x\sqrt{g} \, \xi^{\dagger} \epsilon^{\pm} = 0 = \int_{S^4} d^4x\sqrt{g} \, \rho^{\dagger} \epsilon^{\pm}$. Here, $\epsilon^{\pm}$ are Killing spinors, that is spinors satisfying
    \begin{align}\label{def: Killing spinors very nice}
        \nabla_{\mu}\epsilon^{\mp} \pm \frac{i}{2}\gamma_{\mu} \epsilon^{\mp} = 0 ~; 
    \end{align}
    \item one out of the three spinors $\xi$, $\rho$, $B$ is orthogonal to ``spinor potentials'' $\phi^{\pm}$ for non-isometric conformal Killing vector-spinors (see Appendix \ref{App: conformal Killing vec-spin} for details and explanation of the terminology). We choose $B$, so that $\int_{S^4} d^4x\sqrt{g} \, B^{\dagger} \phi^{\pm} = 0$, where $\phi^{\pm}$ are  solutions of
    \begin{align} \label{def: spinor potentials very nice}
    \left( \nabla_{(\mu}\nabla_{\nu)} \pm i \gamma_{(\mu} \nabla_{\nu)} + \frac{3}{4}g_{\mu \nu}  \right) \phi^{\mp} = 0 ~. 
    \end{align}
\end{itemize}
The requirements in the three bullet points above are a direct consequence of the spherical harmonic analysis in Appendix \ref{appendix: longitudinal modes}. At this point, it is convenient to re-express (\ref{psi decomposition 2}) in a way that makes a pure gauge term appear explicitly:
\begin{align} \label{psi decomposition 3}
    \Psi_{\mu\nu} & = \Psi_{\mu\nu}^{(\text{TT})} + \left( \nabla_{(\mu} \lambda_{\nu)} + \frac{i}{2}\gamma_{(\mu}\lambda_{\nu)} \right) - \frac{i}{2}\gamma_{(\mu}\lambda_{\nu)} + \gamma_{(\mu} \chi_{\nu)} + g_{\mu\nu} B \nonumber \\
    & = \Psi_{\mu\nu}^{(\text{TT})} + \left( \nabla_{(\mu} \lambda_{\nu)} + \frac{i}{2}\gamma_{(\mu}\lambda_{\nu)} \right) + \gamma_{(\mu} {X}_{\nu)} + g_{\mu\nu} B ~,
\end{align}
where we defined the gamma-traceless vector-spinor
\begin{equation} \label{X intro}
    X_\mu \equiv - \frac{i}{2}\lambda_{\mu} + \chi_{\mu} ~, ~~~\text{with}~~~\gamma^\mu X_\mu=0 ~.
\end{equation}
Since the above change of variables is just a shift in $\chi_\mu$, the corresponding Jacobian introduced in the path integral is trivial. We can further split $X_\mu$ into its TT and longitudinal components as
\begin{equation} \label{tilde chi split}
    X_\mu = X_\mu^{(\text{TT})} + \left( \nabla_\mu - \frac{1}{4} \gamma_\mu \slashed{\nabla} \right) P ~,
\end{equation}
where $P$ is a spinor whose Killing spinor modes have been removed, as dictated by the requirement that (\ref{psi decomposition 2}) is not redundant. 
\newline\newline
To sum up, our spin-$\frac{5}{2}$ field is decomposed as%
\footnote{In this decomposition, $\lambda^{(\text{TT})\mu}$, $\xi$, $P$ and $B$ are subject to the restrictions discussed above, which can be summarised by 
\begin{equation} 
    0=\int_{S^4} d^4x\sqrt{g} \, \lambda^{(\text{TT}) \, \mu \, \dagger} A^{\pm}_{\mu} = \int_{S^4} d^4x\sqrt{g} \, \xi^{\dagger} \epsilon^{\pm} =  \int_{S^4} d^4x\sqrt{g} \, P^{\dagger} \epsilon^{\pm} =\int_{S^4} d^4x\sqrt{g} \, B^{\dagger} \phi^{\pm} ~,
\end{equation}
and their Hermitian conjugates.}
\begin{multline} \label{vasilis pretty equation}
    \Psi_{\mu\nu} = \Psi_{\mu\nu}^{(\text{TT})} + \bigl( \nabla_{(\mu} + \frac{i}{2}\gamma_{(\mu} \bigr) \Big(\lambda_{\nu)}^{(\text{TT})} +  \nabla_{\nu)}\xi - \frac{1}{4} \gamma_{\nu)} \slashed{\nabla}  \xi \Big)  + g_{\mu\nu} B \\
    + \gamma_{(\mu} \Big( X_{\nu)}^{(\text{TT})} +  \nabla_{\nu)}P - \frac{1}{4} \gamma_{\nu)} \slashed{\nabla} P \Big) ~.    
\end{multline}
%


\subsection{Action decomposition}

We wish to rewrite the original action (\ref{spin 5/2 action Euclidean}) in terms of the decomposed field (\ref{psi decomposition 3}). Plugging (\ref{psi decomposition 3}) into (\ref{spin 5/2 action Euclidean}) and exploiting the gauge invariance of the action, one is left with
\begin{equation} \label{newly rewritten action}
    S_{E}[\Psi_{\mu\nu}] \equiv S[\Psi_{\mu\nu}^{(\text{TT})}] + S[\gamma_{(\mu} X_{\nu)}] + S[g_{\mu\nu} B] + \text{cross terms} ~.
\end{equation}
The cross terms are 
\begin{equation}\label{action cross terms}
    S[X^\dagger_\mu, B] + S[B^\dagger, X_\mu] = 12 \int_{S^4}~d^4 x \sqrt{g}\left( X^{\dagger \,\mu} \nabla_\mu B + B^\dagger \, \nabla^{\nu}  X_{\nu} \right)~.
\end{equation}
Moreover, we have
\begin{equation}
    S[\Psi_{\mu\nu}^{(\text{TT})}] = \int_{S^4} d^4x~\sqrt{g}~\Psi_{\mu\nu}^{(\text{TT})\dagger}\left( \slashed{\nabla} + 2i \right)\Psi^{(\text{TT})\mu \nu} ~,
\end{equation}
and
\begin{equation} \label{action chi tilde}
    S[\gamma_{(\mu} X_{\nu)}] + S[g_{\mu\nu} B] = \int \left[ 10X^{\mu \, \dagger} (\slashed{\nabla}-3i) X_\mu - 4 B^\dagger (\slashed{\nabla} + 6i) B \right] ~.
\end{equation}
In order to compute the path integral, we further split $X_\mu$ into its TT and longitudinal components, as shown in (\ref{tilde chi split}). Then, $ S_E[\Psi_{\mu\nu}]$ reduces to
\begin{align} \label{final decomposed action} 
    S_E[\Psi_{\mu\nu}] = & \int d^4x \sqrt{g}~\Psi_{\mu\nu}^{(\text{TT})\dagger}\left(   \slashed{\nabla} + 2i\right)\Psi^{(\text{TT})\mu \nu} +10~\int d^4x \sqrt{g}~~X_{\mu}^{(\text{TT}) \dagger}\left(   \slashed{\nabla} - 3i\right)X^{(\text{TT})\mu } \nonumber \\
    & -\frac{15}{4} \int d^4x \sqrt{g}~ P^\dagger (\slashed{\nabla}-6i)(\slashed{\nabla}-2i) (\slashed{\nabla}+2i)P \nonumber \\
    & - 4 \int d^4x \sqrt{g}~~ B^\dagger (\slashed{\nabla} + 6i) B  \nonumber \\
    & - 9 \int d^4x \sqrt{g}~ \Big(P^\dagger (\slashed{\nabla}-2i)(\slashed{\nabla}+2i) B -  B^\dagger(\slashed{\nabla}-2i)(\slashed{\nabla}+2i) P \Big) ~.
\end{align}
The only surviving cross terms now relate two pairs of Dirac spinors, $P^\dagger, B$ and $B^\dagger ,P$, and no longer a spinor and a vector-spinor as in (\ref{action cross terms}). This allows us to compute the Euclidean path integral for (\ref{final decomposed action}) and express it in terms of functional determinants. 

\subsection{Path integral setup and action contribution}

Let us proceed by computing the path integral (\ref{sphere path integral 5/2}) using the explicit form of the action (\ref{final decomposed action}). In what follows, the exclusion of Killing vector-spinor modes (\ref{def: Killing vec-spinors very nice}) from path integration will be denoted by a double prime on the integration measure, as ${\mathcal{D}''}$. The exclusion of spinor modes  that correspond to spinor potentials (\ref{def: spinor potentials very nice})  will be denoted by a tilde on the integration measure, $\tilde{\mathcal{D}}$, and exclusion of Killing spinor modes (\ref{def: Killing spinors very nice}) will be denoted with a bar, as  $\bar{\mathcal{D}}$. This notation will be also employed for the omission of such modes from the product of eigenvalues in spin-$\tfrac{1}{2}$ functional determinants, as $\widetilde{\det}^{(\frac{1}{2})}$ and $\overline{\det}^{(\frac{1}{2})}$ respectively.
\newline\newline
Using (\ref{final decomposed action}), we find that the sphere path integral (\ref{sphere path integral 5/2}) can be re-expressed as%
\footnote{Note that, for instance, $\int \mathcal{D}'' \lambda^{(\text{TT})}$ denotes integration over the full field $\lambda^{(\text{TT})}_{\mu}$ \textit{except} for Killing vector-spinor modes, whereas $\int \mathcal{D} {X}'^{(\text{TT})}$ denotes integration over the ${X}'^{(\text{TT})}_{\mu}$ modes \textit{only} (these modes are defined below (\ref{path integral 1/2 from action})). We know the notation is a bit convoluted and apologise for it in advance, we did our best in trying to simplify it.}%
\begin{align} \label{sphere path integral 5/2 new}
    \mathcal{Z} = \, & \frac{\int \mathcal{D}'' \lambda^{(\text{TT})} \mathcal{D}'' \lambda^{(\text{TT}) \, \dagger} {\bar{\mathcal{D}}}\xi {\bar{\mathcal{D}}}\xi^\dagger \int \mathcal{D} {X}'^{(\text{TT})} \mathcal{D} {X}'^{(\text{TT}) \, \dagger} \int \mathcal{D}P^* \mathcal{D}P^{* \, \dagger} \int \mathcal{D}B^\star \mathcal{D}B^{\star \, \dagger}}{\text{vol} \, \mathcal{G}}  \nonumber \\
    & \times\mathcal{Z}^{(\tfrac{5}{2})}_\Psi \mathcal{Z}^{(\tfrac{3}{2})}_{X} \mathcal{Z}^{(\tfrac{1}{2})}_{P,B} \times J \equiv \mathcal{Z}_{\text{gauge vol}} \times \mathcal{Z}_{\text{action}} \times J ~,
\end{align}
where $J$ is the Jacobian associated with the change of variables (\ref{vasilis pretty equation}). The details regarding $J$ and $\mathcal{Z}_{\text{gauge vol}}$ can be found in subsections \ref{subsec Jacobian} and \ref{subsec gauge volume} respectively. As far as $\mathcal{Z}_{\text{action}}$ is concerned, we have 
%
\begin{align}
  \mathcal{Z}_{\text{action}}= \mathcal{Z}^{(\tfrac{5}{2})}_\Psi   \mathcal{Z}^{(\tfrac{3}{2})}_X  \mathcal{Z}^{(\tfrac{1}{2})}_{P, B} ~,
\end{align}
where
\begin{align}
    \mathcal{Z}^{(\tfrac{5}{2})}_\Psi & \equiv \int \mathcal{D} \Psi^{(\text{TT})} \mathcal{D} \Psi^{(\text{TT}) \, \dagger} e^{-S[\Psi^{(\text{TT})}_{\mu\nu}]} \\
    & = \int \mathcal{D} \Psi^{(\text{TT})} \mathcal{D} \Psi^{(\text{TT}) \, \dagger} \exp \left( - \int \Psi^{(\text{TT}) \,\mu\nu \,\dagger} (\slashed{\nabla}+2i) \Psi_{\mu\nu}^{(\text{TT})} \right) = \text{det}_{\text{TT}}^{(\frac{5}{2})} \left( \frac{\slashed{\nabla}+2i}{\Lambda_{\text{u.v.}}^{\frac{1}{2}}} \right) \nonumber ~,
\end{align}
\begin{align} \label{path integral 3/2 from action}
    \mathcal{Z}^{(\tfrac{3}{2})}_{X} & \equiv \int \mathcal{D}' {X}^{(\text{TT})} \mathcal{D}' {X}^{(\text{TT}) \, \dagger} e^{-S[\gamma_{(\mu} {X}^{(\text{TT})}_{\nu)}]} \\
    & = \int \mathcal{D}' {X}^{(\text{TT})} \mathcal{D}' {X}^{(\text{TT}) \, \dagger} \exp \left( - 10 \int X^{(\text{TT}) \,\mu \,\dagger} (\slashed{\nabla}-3i) X_{\mu}^{(\text{TT})} \right) \propto \text{det}_{\text{TT}}'^{\, (\frac{3}{2})} \left( \frac{\slashed{\nabla}-3i}{\Lambda_{\text{u.v.}}^{\frac{1}{2}}} \right) \nonumber ~,
\end{align}
and
\begin{align} \label{path integral 1/2 from action}
    \mathcal{Z}^{(\tfrac{1}{2})}_{P,B} & \equiv \int \mathcal{\bar{D}}^* P \mathcal{\bar{D}}^* P^\dagger ~\mathcal{\tilde{D}}^{\star} B ~\mathcal{\tilde{D}}^\star B^{\dagger}~ e^{-S[\gamma_{(\mu}(\nabla_{\nu)}-\frac{1}{4}\gamma_{\nu)})P, \, g_{\mu\nu}B]} \\
    & \propto \int \mathcal{\bar{D}}^* P \mathcal{\bar{D}}^* P^\dagger \mathcal{\tilde{D}}^{\star} B \mathcal{\tilde{D}}^\star B^{\dagger} \begin{pmatrix}
        P^\dagger & B^\dagger
    \end{pmatrix} \begin{pmatrix}
        (\slashed{\nabla}-6i)(\slashed{\nabla}^2+4) && (\slashed{\nabla}^2+4) \\
        -(\slashed{\nabla}^2+4) && (\slashed{\nabla}+6i)
    \end{pmatrix} \begin{pmatrix}
        P \\ B
    \end{pmatrix} \nonumber \\
    & = f(\bar{B},B^*,\tilde{P},P^{\star}) \times \widetilde{\overline{\text{det}}}^{* \star \, (\frac{1}{2})} \, \frac{(\slashed{\nabla}^2+4)(\slashed{\nabla}^2+9)}{\Lambda^2_{\text{u.v.}}} \nonumber ~.
\end{align}
In the above, $\Lambda_{\text{u.v.}}$ is some reference high-energy cutoff scale with units of inverse length squared. Whenever we use the symbol $\propto$, the proportionality factor is a constant real number independent of $\Lambda_{\text{u.v.}}$. Moreover:
\begin{itemize}
    \item the prime in $\mathcal{D}' {X}^{(\text{TT})}$ designates the exclusion from path integration in $\mathcal{Z}^{(\tfrac{3}{2})}_{X}$ of vector-spinor modes that satisfy $\slashed{\nabla}{X}'^{(\text{TT})}_\mu = 3i{X}'^{(\text{TT})}_\mu$, that is $n=1$ modes in $\slashed{\nabla}{X}^{(\text{TT})}_\mu = i(n+2){X}^{(\text{TT})}_\mu$ \cite{Letsios:2022tsq, Homma:2020has}. These modes are momentarily excluded from path integration since they are zero modes of the action $S[\gamma_{(\mu} {X}^{(\text{TT})}_{\nu)}]$ appearing in (\ref{path integral 3/2 from action}).  The notation $\text{det}_{\text{TT}}'^{\, (\frac{3}{2})}(\slashed{\nabla} - 3i)$ symbolises a spin-$\tfrac{3}{2}$ functional determinant taken over the full TT eigenvalue spectrum of the corresponding differential operator, except for those eigenvalues corresponding to the zero modes;
    \item  the asterisk in $\mathcal{D}^* P$\footnote{Thus, $\bar{\mathcal{D}}^* P$ denotes the exclusion of both ${P^*}$ modes and all Killing spinors modes $\bar{P}$ (see \eqref{def: Killing spinors very nice}). Similarly, $\tilde{\mathcal{D}}^\star B$ denotes the exclusion of both $B^{\star}$ modes and all spinor potentials $\tilde{B}$ (see \eqref{def: spinor potentials very nice}).} denotes the exclusion from path integration in $\mathcal{Z}^{(\tfrac{1}{2})}_{P,B}$ of spinor modes that satisfy $\slashed{\nabla}P^* = 6iP^*$, that is $n=4$ modes in $\slashed{\nabla}P = i(n+2)P$ \cite{Camporesi:1995fb}. These modes are momentarily excluded from path integration since they are zero modes of the action $S[\gamma_{(\mu}(\nabla_{\nu)}-\frac{1}{4}\gamma_{\nu)})P, \, g_{\mu\nu}B]$ appearing in (\ref{path integral 1/2 from action}); 
    \item the star in $\mathcal{{D}}^\star B$ denotes the exclusion from path integration in $\mathcal{Z}^{(\tfrac{1}{2})}_{P,B}$ of spinor modes that satisfy $\slashed{\nabla}B^{\star} = -6iB^{\star}$, that is $n=4$ modes in $\slashed{\nabla}B = -i(n+2)B$ \cite{Camporesi:1995fb}. Once again, these modes are momentarily excluded from path integration since they are zero modes of the action $S[\gamma_{(\mu}(\nabla_{\nu)}-\frac{1}{4}\gamma_{\nu)})P, \, g_{\mu\nu}B]$ appearing in (\ref{path integral 1/2 from action}); 
    \item the notation $\widetilde{\overline{\text{det}}}^{* \star \, (\frac{1}{2})} \, \Big({(\slashed{\nabla}^2+4)(\slashed{\nabla}^2+9)} \Big)$ in (\ref{path integral 1/2 from action}) symbolises a spin-$\tfrac{1}{2}$ functional determinant taken over the full eigenvalue spectrum of the corresponding differential operator, except for eigenvalues corresponding to four types of modes that have been excluded, denoted by the four symbols $-, \sim, *, \star$. For more details see the discussion below (\ref{explaning det in appendix}). 
\end{itemize}
Note that we must still path integrate over all such modes ($X'^{\text{(TT)}}_{\mu}$, $P^*$, $B^\star$) to compute the full $\mathcal{Z}$, as well as over ${{\lambda}}^{(\text{TT})}_\mu$ and $\xi$, because not doing so would render the problem non-local. Indeed, path integration over these modes is explicitly written down in the first line of (\ref{sphere path integral 5/2 new}) and has been reabsorbed in $\mathcal{Z}_{\text{gauge vol}}$. Further details on the choice of notation and the aforementioned exclusion of modes from path integration can be found in Appendix \ref{Appendix: excluded modes}.
\newline\newline
The function $f(\bar{B},B^*,\tilde{P},P^{\star})$\footnote{Note that, while we are excluding $\bar{P}$, $P^*$, $\tilde{B}$, $B^\star$ modes from path integration in (\ref{path integral 1/2 from action}), there is nothing wrong with $\bar{B}$, $B^*$, $\tilde{P}$, $P^\star$ modes.} appearing in the last line of (\ref{path integral 1/2 from action}) is given by  
\begin{multline}  \label{Chiara's compromise for f(B,rho etc)}
    f(\bar{B},B^*,\tilde{P},P^{\star}) = \int \mathcal{D} {\tilde{P}} \mathcal{D} {\tilde{P}}^{\dagger} \mathcal{D} P^\star \mathcal{D} P^{\star \, \dagger} \exp{\Bigg( \frac{15}{4} \int d^4x \sqrt{g} ~ P^\dagger (\slashed{\nabla}-6i)(\slashed{\nabla}-2i) (\slashed{\nabla}+2i) P \Bigg)} \\
    \times \int \mathcal{D} \bar{B} \mathcal{D} {\bar{B}}^{\dagger} \mathcal{D} B^{*} \mathcal{D} B^{* \, \dagger} \exp{\Bigg( 4 \int d^4x \sqrt{g}~ B^\dagger (\slashed{\nabla} + 6i) B\Bigg)} ~.
\end{multline}
Recall that the Dirac operator acts on spinor spherical harmonics $\psi^{\pm}_n$ on $S^4$ as \cite{Camporesi:1995fb}
\begin{equation} \label{dirac spin 1/2 spectrum}
    \slashed{\nabla} \psi^\pm_n = \pm i \, (n+2) \, \psi^{\pm}_n ~, ~~~\text{with}~~~ n=0,1, \dots ~,
\end{equation}
where the degeneracy of the $n$-th eigenvalue of either sign is given by\footnote{The degeneracy of the eigenvalue $+i(n+2)$ is equal to the degeneracy of the eigenvalue $-i(n+2)$ \cite{Camporesi:1995fb}.}
\begin{equation} \label{degeneracy spin 1/2 spherical harmonics}
    D^{5}_{n,\frac{1}{2}} = \frac{(3+n)!}{n!}\frac{4}{3!} = \frac{2}{3} (n+1)(n+2)(n+3) ~.
\end{equation}
We then find that $f(\bar{B},B^*,\tilde{P},P^{\star}) $ can be expressed in terms of products of eigenvalues of the corresponding differential operators in (\ref{Chiara's compromise for f(B,rho etc)}), as
\begin{multline} \label{extra modes from action VASILIS}
f(\bar{B},B^*,\tilde{P},P^{\star}) = \prod_{\lambda \in \{ 2,-2,6\}}  \left( \frac{4(i \lambda+6i)}{\Lambda_{\text{u.v.}}^{\frac{1}{2}}} \right)^{D_{|\lambda|-2,\frac{1}{2}}^5} \\
    \times \prod_{\gamma \in \{ 3,-3,-6\}}  \left( \frac{15}{4} \frac{(i \gamma - 6i)(-\gamma^2+4)}{\Lambda_{\text{u.v.}}^{\frac{3}{2}}} \right)^{D_{|\gamma|-2,\frac{1}{2}}^5} ~.
\end{multline}
Appendix \ref{Appendix: excluded modes} contains further details on the computation. At first glance, it is not clear whether $f(\bar{B},B^*,\tilde{P},P^{\star})$ is real or complex. If $f(\bar{B},B^*,\tilde{P},P^{\star})$ were complex, this would give rise to a fermionic analogue of Polchinski's phase \cite{Polchinski:1988ua}, which is known to appear in the sphere path integral of strictly and partially massless spin-$s \geq 2$ bosons \cite{Anninos:2020hfj}. However, it is easy to see that $D^{5}_{n,s}$ is always an integer multiple of 4 for generic spin-$s$,\footnote{This can be understood as follows. The $so(5)$ representation furnished by  spin-$s$ TT fermionic spherical harmonics on $S^4$ has $so(5)$ highest weight $(n+\frac{1}{2},s)$ with $n \in \mathbb{N}$ and $n+\frac{1}{2} \geq s$, see \cite{Homma:2020has}. The dimension of this representation is $D^5_{n,s} = \frac{1}{3} (n+2)(2s+1)(n-s+\frac{3}{2})(n+s+\frac{5}{2})$. If we express the spin as $s=\frac{2m+1}{2}$, it's easy to see that when $m$ is odd, $(2s+1)$ is an integer multiple of 4 for any $n$. When $m$ is even, if $n$ is also even then $(2s+1)(n+2)$ is an integer multiple of 4, whereas if $n$ is odd then $(2s+1)(n-s+\frac{3}{2})$ is an integer multiple of 4.} and thus $f(\bar{B},B^*,\tilde{P},P^{\star})$ is real, which means that there is  no analogue of Polchinski's phase for the strictly massless spin-$\frac{5}{2}$ field. In \cite{Anninos:2025mje}, we have also shown that no such phase appears in the $S^4$ path integral of the strictly massless spin-$\tfrac{3}{2}$ field. We expect that no such phase will appear for any strictly massless fermion of generic spin-$s$ on $S^4$. \\\\ 
A direct evaluation of (\ref{extra modes from action VASILIS}) yields
\begin{equation} f(\bar{B},B^*,\tilde{P},P^{\star}) \propto \left( \Lambda_{\text{u.v.}} \right)^{-332} ~,
\end{equation}
where the proportionality constant is a real number whose explicit value is not relevant for our analysis.


\subsection{Jacobian} \label{subsec Jacobian}

As our local measure of path integration, we select the following
\begin{equation} \label{eq: psi measure}
    1 = \int \mathcal{D} \Psi \mathcal{D} \Psi^\dagger \exp \left( - \Lambda_{\text{u.v.}}^{\frac{1}{2}} \int d^4 x \sqrt{g} \, g^{\mu\nu} g^{\rho\sigma} \Psi^\dagger_{\mu\rho} \Psi_{\nu\sigma} \right) ~.   
\end{equation}
When introducing the change of variables (\ref{psi decomposition 2})-(\ref{chi decomposition}) (or equivalently (\ref{vasilis pretty equation}), since (\ref{X intro}) is just a shift), we must preserve the above normalisation condition and therefore introduce a Jacobian factor. In other words, we need a factor $J$ such that
\begin{align} \label{eq: psi expanded measure}
    1 = & \int \mathcal{D} \Psi \mathcal{D} \Psi^\dagger \exp \left( - \Lambda_{\text{u.v.}}^{\frac{1}{2}} \int d^4 x \sqrt{g} \, g^{\mu\nu} g^{\rho\sigma} \Psi^\dagger_{\mu\rho} \Psi_{\nu\sigma} \right)  \nonumber \\  
     =& \int \mathcal{D} \Psi^{(\text{TT})} \mathcal{D} \Psi^{(\text{TT}) \, \dagger} \bar{\mathcal{D}}'' \lambda \bar{\mathcal{D}}'' \lambda^\dagger \bar{\mathcal{D}} {\chi} \bar{\mathcal{D}} {\chi}^\dagger \tilde{\mathcal{D}} B \tilde{\mathcal{D}} B^\dagger \times J \times \exp \Biggl( - \Lambda_{\text{u.v.}}^{\frac{1}{2}} \int d^4 x \sqrt{g} \, g^{\mu\nu} g^{\rho\sigma} \nonumber \\
    & \biggl[\Psi_{\mu\rho}^{(\text{TT})} + \nabla_{(\mu} \lambda_{\rho)} + \gamma_{(\mu} {\chi_{\rho)}} + g_{\mu\rho}B \biggr]^\dagger \biggl[ \Psi_{\nu\sigma}^{(\text{TT})} + \nabla_{(\nu} \lambda_{\sigma)} + \gamma_{(\nu} {\chi_{\sigma)}} + g_{\nu\sigma}B \biggr] \Biggr) ~.
\end{align}
Here, the vector-spinors $\lambda_\mu$ and $\chi_\mu$ are understood to be decomposed in their TT and longitudinal parts, as%
\footnote{Recall that the spinors $\xi$ and $\rho$ are taken to be orthogonal to Killing spinors (\ref{def: Killing spinors very nice}).} %
\begin{equation}
    \lambda_\mu = \lambda_\mu^{(\text{TT})} + \left( \nabla_\mu - \frac{1}{4} \gamma_\mu \slashed{\nabla} \right) \xi ~,
\end{equation}
\begin{equation}
    {\chi}_\mu = {\chi}_\mu^{(\text{TT})} + \left( \nabla_\mu - \frac{1}{4} \gamma_\mu \slashed{\nabla} \right) \rho ~.
\end{equation}
We also define $\bar{\mathcal{D}}''\lambda \equiv {\mathcal{D}}''\lambda^{(\text{TT})} {\bar{\mathcal{D}}}\xi$ and $\bar{\mathcal{D}}\chi \equiv {\mathcal{D}}\chi^{(\text{TT})} {\bar{\mathcal{D}}}\rho$, to have a more compact notation. The argument of the exponential in (\ref{eq: psi expanded measure}) can be conveniently expressed as
\begin{multline} \label{eq: second expanded exponential argument}
     g^{\mu\nu} g^{\rho\sigma} \biggl[\Psi_{\mu\rho}^{(\text{TT})} + \nabla_{(\mu} \lambda_{\rho)} + \gamma_{(\mu} {\chi_{\rho)}} + g_{\mu\rho}B \biggr]^\dagger \biggl[ \Psi_{\nu\sigma}^{(\text{TT})} + \nabla_{(\nu} \lambda_{\sigma)} + \gamma_{(\nu} {\chi_{\sigma)}} + g_{\nu\sigma}B \biggr] \\
     = \Psi^{(\text{TT}) \,\mu\nu \,\dagger} \Psi_{\mu\nu}^{(\text{TT})} + A_\frac{3}{2} + A_\frac{1}{2} ~,
\end{multline}
where
\begin{equation}
    A_\frac{3}{2} = \begin{pmatrix}
        \lambda_\mu^{(\text{TT}) \, \dagger} & \chi_\mu^{(\text{TT}) \, \dagger}
    \end{pmatrix} \begin{pmatrix}
        -\frac{1}{2}(\slashed{\nabla}^2+\frac{15}{2}) && -\frac{1}{2}\slashed{\nabla} \\\\
        \frac{1}{2}\slashed{\nabla} && 3
    \end{pmatrix} \begin{pmatrix}
        \lambda^{(\text{TT}) \, \mu} \\\\ \chi^{(\text{TT}) \, \mu}
    \end{pmatrix} ~,
\end{equation}
and
\begin{equation} \label{spin 1/2 jacobian A}
    A_\frac{1}{2} = \begin{pmatrix}
        \xi^\dagger & \rho^\dagger & B^\dagger
    \end{pmatrix} \begin{pmatrix}
        \frac{3}{32}(\slashed{\nabla}^2+4)(7\slashed{\nabla}^2+54) && \frac{3}{16}\slashed{\nabla}(\slashed{\nabla}^2+4) && \frac{3}{4}(\slashed{\nabla}^2+4) \\\\
        -\frac{3}{16}\slashed{\nabla}(\slashed{\nabla}^2+4) && -\frac{9}{4}(\slashed{\nabla}^2+4) && 0 \\\\
        \frac{3}{4}(\slashed{\nabla}^2+4) && 0 && 4
    \end{pmatrix} \begin{pmatrix}
        \xi \\\\ \rho \\\\ B
    \end{pmatrix} ~.
\end{equation}
Thus, (\ref{eq: psi expanded measure}) simplifies to
\begin{align} \label{eq: new expanded measure}
    1 = & \int \mathcal{D} \Psi^{(\text{TT})} \mathcal{D} \Psi^{(\text{TT}) \, \dagger} \mathcal{D}'' \lambda^{(\text{TT})} \mathcal{D}'' \lambda^{(\text{TT}) \, \dagger} {\bar{\mathcal{D}}}\xi {\bar{\mathcal{D}}}\xi^\dagger \mathcal{D} {\chi}^{(\text{TT})} \mathcal{D} {\chi}^{(\text{TT}) \, \dagger} \bar{\mathcal{D}} \rho \bar{\mathcal{D}} \rho^\dagger \tilde{\mathcal{D}} B \tilde{\mathcal{D}} B^\dagger \nonumber \\ 
    & J \times \exp \left( - \Lambda_{\text{u.v.}}^{\frac{1}{2}} \int d^4 x \sqrt{g} \left[ \Psi^{(\text{TT}) \,\mu\nu \,\dagger} \Psi_{\mu\nu}^{(\text{TT})} + A_\frac{3}{2} + A_\frac{1}{2} \right] \right)~.
\end{align}
Defining the measures governing the fields in the decomposition of $\Psi_{\mu\nu}$ as
\begin{eqnarray} \label{eq: measures}
    1 &\equiv& \int \mathcal{D} \Psi^{(\text{TT})} \mathcal{D} \Psi^{(\text{TT}) \, \dagger} \exp \left( - \Lambda_{\text{u.v.}}^{\frac{1}{2}} \int d^4 x \sqrt{g} \, \Psi^{(\text{TT}) \,\mu\nu \,\dagger} \Psi_{\mu\nu}^{(\text{TT})} \right) ~,   \\ 
    1 &\equiv& \int \mathcal{D} \lambda^{(\text{TT})} \mathcal{D} \lambda^{(\text{TT}) \, \dagger}  \exp \left( - \Lambda_{\text{u.v.}}^{\frac{3}{2}} \int d^4 x \sqrt{g} \, \lambda^{(\text{TT}) \, \mu \, \dagger} \lambda_\mu^{(\text{TT})} \right) ~,  \\
    1 &\equiv& \int \mathcal{D} \chi^{(\text{TT})} \mathcal{D} \chi^{(\text{TT}) \, \dagger}  \exp \left( - \Lambda_{\text{u.v.}}^{\frac{1}{2}} \int d^4 x \sqrt{g} \, \chi^{(\text{TT}) \, \mu \, \dagger} \chi_\mu^{(\text{TT})} \right) ~,   \\
    1 &\equiv& \int \mathcal{D} \xi \mathcal{D} \xi^\dagger \exp \left( - \Lambda_{\text{u.v.}}^{\frac{5}{2}} \int d^4 x \sqrt{g} \, \xi^{\dagger} \xi \right) ~, \\
    1 &\equiv& \int \mathcal{D} \rho \mathcal{D} \rho^\dagger \exp \left( - \Lambda_{\text{u.v.}}^{\frac{3}{2}} \int d^4 x \sqrt{g} \, \rho^{\dagger} \rho \right) ~, \\
    1 &\equiv& \int \mathcal{D} B \mathcal{D} B^\dagger \exp \left( - \Lambda_{\text{u.v.}}^{\frac{1}{2}} \int d^4 x \sqrt{g} \, B^{\dagger} B \right) ~,
\end{eqnarray}
we readily find that the Jacobian can be expressed in terms of functional determinants. In particular,
\begin{align}
    J = \tfrac{1}{Y_{\lambda,\chi}^{(\frac{3}{2})} Y_{\xi,\rho,B}^{(\frac{1}{2})}} ~,
\end{align}
where
\begin{equation}
    Y_{\lambda,\chi}^{(\frac{3}{2})} = \text{local} \times \text{det}_{\text{TT}}''^{\, (\frac{3}{2})} \frac{5(-\slashed{\nabla}^2-9)}{4\Lambda_{\text{u.v.}}} \propto \text{det}_{\text{TT}}''^{\, (\frac{3}{2})}   \frac{\slashed{\nabla}^2+9}{\Lambda_{\text{u.v.}}} ~, \label{eq: Jacobian det 3/2}
\end{equation}
\begin{align}
    Y_{\xi,\rho,B}^{(\frac{1}{2})} = & ~ \text{local} \times \prod_{\gamma \in \{ 3,-3\}}  \left( \frac{9}{128} \frac{(\frac{41}{2}\gamma^2 - 162)(-\gamma^2+4)^2}{\Lambda_{\text{u.v.}}^{3}} \right)^{D_{|\gamma|-2,\frac{1}{2}}^5} \nonumber \\ 
    & \times \widetilde{\overline{\text{det}}}^{(\frac{1}{2})} \frac{-9(\slashed{\nabla}^2+4)^2 (\slashed{\nabla}^2+9)}{2\Lambda^3_{\text{u.v.}}} \propto \left( \Lambda_{\text{u.v.}} \right)^{-96} \times \widetilde{\overline{\text{det}}}^{(\frac{1}{2})} \frac{(\slashed{\nabla}^2+4)^2 (\slashed{\nabla}^2+9)}{\Lambda^3_{\text{u.v.}}} \label{eq: Jacobian det 1/2} ~.
\end{align}
In the above, the term `local' refers to functional determinants taken over constant operators which are independent of the cutoff $\Lambda_{\text{u.v.}}$.\footnote{In particular, the term `local' in (\ref{eq: Jacobian det 3/2}) stands for 
$$ \int D \chi''^{(\text{TT})} ~D \chi''^{(\text{TT})\dagger}~\exp\Big({- 3 \Lambda_{\text{u.v.}}^{\frac{1}{2}} \int d^4x \sqrt{g}~\chi_\mu^{(\text{TT})\dagger}}~ \chi^{(\text{TT})\mu} \Big) ~,$$
which is independent of the cutoff $\Lambda_{\text{u.v.}}$ because it corresponds to a product of eigenvalues of the operator `$3$' acting on TT vector-spinors. Similarly, the term `local' in (\ref{eq: Jacobian det 1/2}) stands for%
$$ \int D \bar{B} ~ D \bar{B}^{\dagger}~\exp\Big({-4\Lambda_{\text{u.v.}}^{\frac{1}{2}} \int d^4x \sqrt{g}~B^\dagger B} \Big) ~.$$} The product in the first line of (\ref{eq: Jacobian det 1/2}) appears for the same reason as (\ref{extra modes from action VASILIS}).

\subsection{Residual gauge group volume} \label{subsec gauge volume}

The factor $\mathcal{Z}_{\text{gauge vol}}$ in (\ref{sphere path integral 5/2 new}), which encodes the gauge group volume contribution to the path integral, is given by
\begin{equation}\label{define Z_gauge vol}
  \mathcal{Z}_{\text{gauge vol}} =  \frac{\int \mathcal{D}'' \lambda^{(\text{TT})} \mathcal{D}'' \lambda^{(\text{TT}) \, \dagger} {\bar{\mathcal{D}}}\xi {\bar{\mathcal{D}}}\xi^\dagger \int \mathcal{D} {X}'^{(\text{TT})} \mathcal{D} {X}'^{(\text{TT}) \, \dagger} \int \mathcal{D}P^* \mathcal{D}P^{* \, \dagger} 
    \int \mathcal{D}B^\star \mathcal{D}B^{\star \, \dagger} } {\text{vol}\,\mathcal{G}} ~.
\end{equation}
In the above, we have accounted for integration over the vector-spinor $\lambda_\mu$ (\ref{vasilis pretty equation}), which drops out of the field decomposition (\ref{newly rewritten action}) because of the gauge invariance of the action, and for the extra modes we had previously excluded from path integration in (\ref{path integral 3/2 from action}) and (\ref{path integral 1/2 from action}), i.e. the zero modes of the spin-$\tfrac{3}{2}$ and spin-$\tfrac{1}{2}$ determinants. Since the Euclidean action (\ref{spin 5/2 action Euclidean}) is invariant under $\Psi_{\mu\nu} \rightarrow \Psi_{\mu\nu} + \left(\nabla_{(\mu} + \frac{i}{2}\gamma_{(\mu} \right)\epsilon_{{\nu})}$, where $\epsilon_{{\nu}}$ is a gamma-traceless vector-spinor, the corresponding gauge group `volume' is 
\begin{equation}
    \text{vol} \, \mathcal{G}  = \int \mathcal{D} \epsilon \mathcal{D} \epsilon^\dagger ~.
\end{equation}
That is, the gauge group volume is given by path integrating over gamma-traceless vector-spinor gauge parameters. It stems from the following locally defined metric for $\epsilon_\mu$,
\begin{equation} \label{eq: metric for gauge parameter}
    ds^2_\epsilon = \Lambda^{\frac{3}{2}}_{\text{u.v.}} \int d^4x \sqrt{g} \, \delta\epsilon^{\mu \, \dagger} \delta\epsilon_\mu ~,
\end{equation}
and it cancels almost perfectly with $\mathcal{D}'' \lambda^{(\text{TT})} \mathcal{D}'' \lambda^{(\text{TT}) \, \dagger} \bar{\mathcal{D}}\xi \bar{\mathcal{D}}\xi^\dagger$ in (\ref{define Z_gauge vol}). In order to make this clear, it is convenient to further decompose the gamma-traceless gauge parameter as
\begin{equation} \label{gauge parameter split}
    \epsilon_\mu = \epsilon_\mu^{(\text{TT})} + \left( \nabla_\mu - \frac{1}{4}\gamma_\mu \slashed{\nabla} \right) \omega ~,
\end{equation}
so that\footnote{Note that if we expand $\omega$ in spherical harmonics as $\displaystyle \omega = \sum_n \sum_{\sigma= \pm} c^\sigma_n \psi^{\sigma}_n$, it is easy to see that the $n=0$ modes $\psi^{\pm}_0$, which are Killing spinors, drop out of (\ref{gauge parameter split}) since $(\nabla_{\mu} -\frac{1}{4}\gamma_\mu \slashed{\nabla} ) \psi^{\pm}_0 =0$. Nevertheless, we still need to path integrate over such modes in (\ref{define volG with new variables}).}
\begin{equation}\label{define volG with new variables}
    \text{vol} \, \mathcal{G}  = \int \mathcal{D} \epsilon^{(\text{TT})} \mathcal{D} \epsilon^{(\text{TT}) \, \dagger} \mathcal{D}\omega \mathcal{D}\omega^\dagger \, J^{(\frac{1}{2})}_{\text{vol}} ~,
\end{equation}
where the split (\ref{gauge parameter split}) has introduced the Jacobian factor $J^{(\frac{1}{2})}_{\text{vol}}$. Demanding that the normalisation condition
\begin{align}
  1 = \int \mathcal{D} \epsilon \mathcal{D} \epsilon^\dagger  \exp \left( - \Lambda_{\text{u.v.}}^{\frac{3}{2}} \int d^4 x \sqrt{g} \, \epsilon^{\mu  \dagger} \epsilon_\mu \right) ~,
\end{align}
is preserved under the change of variables (\ref{gauge parameter split}), and working as in section \ref{subsec Jacobian}, we find%
\begin{equation} \label{J 1/2 expression}
    J_{\text{vol}}^{(\frac{1}{2})} = \frac{1}{\int \mathcal{D}\bar{\omega} \mathcal{D}\bar{\omega}^{\dagger} \times {\overline{\text{det}}}^{(\frac{1}{2})} \left( -\frac{3}{4} \frac{\slashed{\nabla}^2+4}{\Lambda_{\text{u.v.}}} \right)} \propto \frac{\Lambda_{\text{u.v.}}^{20}}{\overline{\text{det}}^{(\frac{1}{2})} \left( \frac{\slashed{\nabla}^2+4}{\Lambda_{\text{u.v.}}} \right)} ~,
\end{equation}
where we extracted the $\Lambda_{\text{u.v}}$ dependence from $\int \mathcal{D}\bar{\omega} \mathcal{D}\bar{\omega}^{\dagger}$.%
\footnote{The power of the cutoff in (\ref{J 1/2 expression}), $\Lambda_{\text{u.v.}}^{20}$, is explained as follows. Consider the  induced measure for the field $\omega$, namely
\begin{equation}
    1 \equiv \int \mathcal{D} \omega \mathcal{D} \omega^\dagger  \exp \left( - \Lambda_{\text{u.v.}}^{\frac{5}{2}} \int d^4 x \sqrt{g} \, \omega^{\dagger} \omega \right) ~.
\end{equation}
The locally defined metric for $\omega$ is
\begin{equation} 
    ds^2_\omega = \Lambda^{\frac{5}{2}}_{\text{u.v.}} \int d^4x \sqrt{g} \, \delta\omega^{  \dagger}\, \delta\omega ~.
\end{equation}
We expand $\omega$ in spinor spherical harmonics as $\omega(x) = \displaystyle \sum_i c_i \psi_{(i)}(x)$, where we collectively denote all quantum numbers with the subscript $_{(i)}$. Because of their normalisation, i.e. $\delta_{ij} = \int_{S^4} d^4x \sqrt{g} ~ \psi_{(j)}(x)^{\dagger} \psi_{(i)}(x)$, the modes $\psi_{(i)}(x)$ have units of inverse length squared, $\sim [\ell^{-2}]$, and consequently the Grassmann-odd coefficients $c_i$ are $\sim [\ell^{\frac{5}{2}}]$ because $\omega$ has units of $[\ell^{\frac{1}{2}}]$. The corresponding dimensionless mode measure is
\begin{equation} \label{def: mode measure for omega}
    ds^2_{\text{mode},\,\omega} = \Lambda^{-\frac{5}{2}}_{\text{u.v.}}  ~\sum_{i} d\bar{c}_{i} dc_{i} ~,
\end{equation}
since the Grassmann-odd measure $dc_{i}$ is $\sim [\ell^{-\frac{5}{2}}]$, i.e. it has inverse units with respect to $c_i$. Then, we find $\int \mathcal{D}\bar{\omega} \mathcal{D}\bar{\omega}^{\dagger} \propto (\Lambda_{\text{u.v.}}^{-\frac{5}{4}})^{16}$, where 16 is the real dimension of the space of Killing spinors on $S^4$ (remember that there exist 4 complex Killing spinor modes such that $\slashed{\nabla}\bar{\omega} = +2i\,\bar{\omega}$ and 4 more such that $\slashed{\nabla}\bar{\omega} = -2i\,\bar{\omega}$, see (\ref{degeneracy spin 1/2 spherical harmonics})). This explains the factor of $\Lambda_{\text{u.v.}}^{20}$ in the numerator of (\ref{J 1/2 expression}).}
Now that we computed $J^{(\frac{1}{2})}_{\text{vol}}$, we are ready to find the gauge group volume contribution to the path integral. In particular, we will split $\mathcal{Z}_{\text{gauge vol}}$ into two distinct terms depending on the spin of the fields being integrated over, namely
\begin{equation} \label{eq: residual volume}
     \mathcal{Z}_{\text{gauge vol}} = \text{res}_{\frac{3}{2}} \times \text{res}_{\frac{1}{2}} ~.
\end{equation}
To begin with, let us examine the spin-$\tfrac{3}{2}$ contribution:
\begin{align} \label{eq: residual volume 3/2}
    \text{res}_{\frac{3}{2}} & = \frac{\int \mathcal{D}'' \lambda^{(\text{TT})} \mathcal{D}'' \lambda^{(\text{TT}) \, \dagger} \int \mathcal{D} {X}'^{(\text{TT})} \mathcal{D} {X}'^{(\text{TT}) \, \dagger} } {\int \mathcal{D} \epsilon^{(\text{TT})} \mathcal{D} \epsilon^{(\text{TT}) \, \dagger}} \nonumber \\
    & = \frac{\int \mathcal{D}'' \lambda^{(\text{TT})} \mathcal{D}'' \lambda^{(\text{TT}) \, \dagger} \int (\sqrt{\Lambda_{\text{u.v.}}})^{20}~\mathcal{D}  {\lambda}'^{(\text{TT})} ~(\sqrt{\Lambda_{\text{u.v.}}})^{20}\mathcal{D} {\lambda}'^{(\text{TT}) \, \dagger} } {\int \mathcal{D} \epsilon^{(\text{TT})} \mathcal{D} \epsilon^{(\text{TT}) \, \dagger}} \\
    & = \frac{(\sqrt{\Lambda_{\text{u.v.}}})^{40}} {\int \mathcal{D} {\epsilon}_-'^{(\text{TT})} \mathcal{D} {\epsilon}_-'^{(\text{TT}) \, \dagger}} = \frac{\Lambda_{\text{u.v.}}^{20}}{\text{vol} \, \mathcal{G}_{\text{KVS}_-}} \propto \Lambda_{\text{u.v.}}^{50} \nonumber ~.
\end{align}
%
In (\ref{eq: residual volume 3/2}), the quantity
\begin{align} \label{vol generated by KVS-}
    \text{vol} \, \mathcal{G}_{\text{KVS}_-} \equiv \int \mathcal{D} {\epsilon}_-'^{(\text{TT})} \mathcal{D} {\epsilon}_-'^{(\text{TT}) \, \dagger} ~,   
\end{align}
denotes the volume of the residual gauge transformations, i.e. `trivial' gauge transformations generated by Killing vector-spinor parameters ${\epsilon}_{- \, \nu} '^{(\text{TT})} $ satisfying $\left(\nabla_{(\mu} + \frac{i}{2} \gamma_{(\mu}  \right) {\epsilon}_{- \, \nu)}'^{(\text{TT})} =0$. \\\\
%
%
Now let us analyse the spin-$\tfrac{1}{2}$ contribution to (\ref{eq: residual volume}):
\begin{align} \label{eq: residual volume 1/2}
    \text{res}_{\frac{1}{2}} & = \frac{\int {\bar{\mathcal{D}}}\xi {\bar{\mathcal{D}}}\xi^\dagger \int \mathcal{D}P^* \mathcal{D}P^{* \, \dagger} 
    \int \mathcal{D}B^\star \mathcal{D}B^{\star \, \dagger}} { J_{\text{vol}}^{(\frac{1}{2})} \int \mathcal{D}\omega \mathcal{D}\omega^\dagger} \nonumber \\
    & = \frac{\int \mathcal{D}P^* \mathcal{D}P^{* \, \dagger} \int \mathcal{D}B^\star \mathcal{D}B^{\star \, \dagger}} { J_{\text{vol}}^{(\frac{1}{2})} \int \mathcal{D}\bar{\omega} \mathcal{D}\bar{\omega}^{\dagger}} \\
    & \propto \frac{1}{J_{\text{vol}}^{(\frac{1}{2})}} \frac{(\Lambda_{\text{u.v.}}^{-\frac{3}{4}})^{280} (\Lambda_{\text{u.v.}}^{-\frac{1}{4}})^{280}}{(\Lambda_{\text{u.v.}}^{-\frac{5}{4}})^{16}} \propto \Lambda_{\text{u.v.}}^{-280} \times {\overline{\text{det}}}^{(\frac{1}{2})} \left( \frac{\slashed{\nabla}^2+4}{\Lambda_{\text{u.v.}}} \right) \nonumber ~.
\end{align}
In conclusion, the contribution to the full path integral coming from the residual gauge group volume amounts to
\begin{equation}\label{final expression for Z_gauge vol}
    \mathcal{Z}_{\text{gauge vol}} = \text{res}_{\frac{3}{2}} \times \text{res}_{\frac{1}{2}} \propto \Lambda_{\text{u.v.}}^{-230} \times {\overline{\text{det}}}^{(\frac{1}{2})} \left( \frac{\slashed{\nabla}^2+4}{\Lambda_{\text{u.v.}}} \right) ~.
\end{equation}

\subsection{Final result and logarithmic divergence coefficient}

Let us finally assemble all our results:
\begin{multline} \label{Z action}
    \mathcal{Z}_{\text{action}} = b \, \Lambda_{\text{u.v.}}^{-332} \times \text{det}_{\text{TT}}^{(\frac{5}{2})} \left( \frac{\slashed{\nabla}+2i}{\Lambda_{\text{u.v.}}^{\frac{1}{2}}} \right) \times \text{det}_{\text{TT}}'^{\, (\frac{3}{2})} \left( \frac{\slashed{\nabla}-3i}{\Lambda_{\text{u.v.}}^{\frac{1}{2}}} \right) \\
    \times \widetilde{\overline{\text{det}}}^{* \star \, (\frac{1}{2})} \, \frac{(\slashed{\nabla}^2+4)(\slashed{\nabla}^2+9)}{\Lambda^2_{\text{u.v.}}} ~,
\end{multline}
\begin{equation} \label{Z Jacobian}
    J = c \, \Lambda_{\text{u.v.}}^{96} \times \left[ \text{det}_{\text{TT}}''^{\, (\frac{3}{2})} \left( \frac{\slashed{\nabla}^2+9}{\Lambda_{\text{u.v.}}} \right)
    \times \widetilde{\overline{\text{det}}}^{(\frac{1}{2})} \, \frac{(\slashed{\nabla}^2+4)^2 (\slashed{\nabla}^2+9)}{\Lambda^3_{\text{u.v.}}} \right]^{-1} ~,
\end{equation}
\begin{equation} \label{Z gauge vol}
    \mathcal{Z}_{\text{gauge vol}} = d \, \Lambda_{\text{u.v.}}^{-230} \times {\overline{\text{det}}}^{(\frac{1}{2})} \, \frac{(\slashed{\nabla}^2+4)}{\Lambda_{\text{u.v.}}} ~.
\end{equation}
Then, using (\ref{sphere path integral 5/2 new}) we obtain
\begin{equation} \label{final spin-5/2 action path integral}
    \mathcal{Z} = \mathcal{Z}_{\text{action}} \mathcal{Z}_{\text{gauge vol}} J = a \, \Lambda_{\text{u.v.}}^{52} \times \frac{\text{det}_{\text{TT}}^{(\frac{5}{2})} \left( \frac{\slashed{\nabla}+2i}{\Lambda_{\text{u.v.}}^{\frac{1}{2}}} \right)}{\text{det}_{\text{TT}}''^{\, (\frac{3}{2})} \left( \frac{\slashed{\nabla}+3i}{\Lambda_{\text{u.v.}}^{\frac{1}{2}}} \right)} ~,
\end{equation}
where overall constants have been reabsorbed in the dimensionless pre-factors $a,b,c,d$. The power of the cutoff,
\begin{equation} \label{power of cutoff}
    52 = -466-10-32+560 ~,
\end{equation}
is explained as follows: $-466$ comes from adding up the explicit powers of $\Lambda_{\text{u.v.}}$ in (\ref{Z action})-(\ref{Z gauge vol}), $-10$ comes from the simplification of the spin-$\tfrac{3}{2}$ sector,\footnote{By simplification of the spin-$\tfrac{3}{2}$ sector, we mean the ratio of determinants $$\frac{\text{det}_{\text{TT}}'^{\, (\frac{3}{2})} \left( \frac{\slashed{\nabla}-3i}{\Lambda_{\text{u.v.}}^{\frac{1}{2}}} \right)}{\text{det}_{\text{TT}}''^{\, (\frac{3}{2})} \left( \frac{\slashed{\nabla}^2+9}{\Lambda_{\text{u.v.}}} \right)} \propto \frac{\Lambda_{\text{u.v.}}^{-10}}{\text{det}_{\text{TT}}''^{\, (\frac{3}{2})} \left( \frac{\slashed{\nabla}+3i}{\Lambda_{\text{u.v.}}^{\frac{1}{2}}} \right)}~,$$ which appears when we take the product of $\mathcal{Z}_{\text{action}}$ and $J$ in (\ref{final spin-5/2 action path integral}). A similar analysis applies to the simplification of the spin-$\tfrac{1}{2}$ sector.} while
$-32$ and $+560$ come from the simplification of the spin-$\tfrac{1}{2}$ sector. In particular, $-32$ stems from the $ \sim$  modes, and  $+560$ comes from the $\star$ and $*$ modes. %
\newline\newline
We will now proceed to study the functional determinants appearing in (\ref{final spin-5/2 action path integral}) to extract the coefficient of the logarithmic UV divergence of the sphere path integral. The reason why we care about computing the exact coefficient of the logarithmic divergence is that such a term is universal, and it cannot be reabsorbed by local counterterms, unlike the $\sim \frac{1}{t^2}$ and $\sim \frac{1}{t^4}$ divergent pieces of $\mathcal{Z}$. Let us begin with the spin-$\tfrac{5}{2}$ determinant in (\ref{final spin-5/2 action path integral}). Employing the methods of \cite{Anninos:2020hfj}, one finds 
\begin{equation}
    \log \text{ det}_{\text{TT}}^{(\frac{5}{2})} \left( \frac{\slashed{\nabla}+2i}{\Lambda_{\text{u.v.}}^{\frac{1}{2}}} \right) = - \int_\varepsilon^\infty \frac{dt}{\sqrt{t^2-\varepsilon^2}} \, 4e^{-4t} \frac{14-16e^{-t}+5e^{-2t}}{(1-e^{-t})^4} \left( e^{2\sqrt{t^2-\varepsilon^2}} + e^{-2\sqrt{t^2-\varepsilon^2}} \right) ~.
\end{equation}
Here, the cutoff parameter $\varepsilon$ is related to $\Lambda_{\text{u.v.}}$ as $\Lambda_{\text{u.v.}} \propto
\varepsilon^{-2}$.   
If we formally set $\varepsilon=0$, taking the small-$t$ expansion of the integrand we find the $\frac{1}{t}$ coefficient
\begin{equation}
    \mathcal{B}_1 = -\frac{571}{30} ~,
\end{equation}
which is associated with the logarithmic divergence. Similarly, for the spin-$\tfrac{3}{2}$ determinant in (\ref{final spin-5/2 action path integral}), we find
\begin{multline}
    \log \text{ det}_{\text{TT}}''^{\,(\frac{3}{2})} \left( \frac{\slashed{\nabla}-3i}{\Lambda^{\frac{1}{2}}_{\text{u.v.}}} \right) = \\
    = - \int_\varepsilon^\infty \frac{dt}{\sqrt{t^2-\varepsilon^2}} \, 4e^{-4t} \frac{16-29e^{-t}+20e^{-2t}-5e^{-3t}}{(1-e^{-t})^4} \left( e^{3\sqrt{t^2-\varepsilon^2}} + e^{-3\sqrt{t^2-\varepsilon^2}} \right) ~.
\end{multline}
If we again set $\varepsilon=0$ and take the small-$t$ expansion of the integrand, we obtain the $\frac{1}{t}$ coefficient
\begin{equation}
    \mathcal{B}_2 = \frac{1129}{45} ~.
\end{equation}
It is then possible to find the overall coefficient of the logarithmic divergent term in the path integral $\mathcal{Z}$ as
\begin{equation} \label{eq: coeff log div func dets}
    \mathcal{B}_{S^4;\frac{5}{2}} = \left( -\frac{571}{30} \right) - \left( \frac{1129}{45} \right) +{52\times2} = \frac{5389}{90} ~,
\end{equation}
where the coefficient $\mathcal{B}_{\text{vol}} \equiv  {+52\times2}$ is due to a proper treatment of the path integration measure -- it stems from the $\Lambda_{\text{u.v.}}$ dependent prefactor in (\ref{final spin-5/2 action path integral}). \\\\
As a side note, it would be interesting to compute Euclidean path integrals and corresponding logarithmic divergences for higher-spin fields living on the subleading saddles (after the sphere) of $\Lambda>0$ Einstein gravity in four dimensions. This work was initiated in \cite{Anninos:2025ltd} for the case of $\mathbb{C}P^2$.


\section{Character integral transform for strictly massless fermions of any spin} \label{spin-s section}

In this section, we will compute the Euclidean path integral on an $S^4$ background for generic spin-$s \geq \tfrac{3}{2}$ totally symmetric fermionic gauge fields.\footnote{The $S^4$ path integral for the strictly massless spin-$\tfrac{3}{2}$ field (gravitino) was first studied in \cite{Anninos:2025mje}.} Unlike our analysis for the spin-$\tfrac{5}{2}$ field above (and the spin-$\tfrac{3}{2}$ field in \cite{Anninos:2025mje}), we will not derive explicitly the form of the path integral in terms of functional determinants. We will instead employ the methods first introduced in \cite{Anninos:2020hfj}, where the authors computed the sphere path integral  for  massive and massless bosonic higher-spin fields. In this section, we will begin with a conjecture for the functional determinant form of the $S^4$ path integral for spin-$s$ fermionic gauge fields (for $s =\tfrac{3}{2}$ and $s =\tfrac{5}{2}$ the conjecture has been verified). We will then proceed to express the $S^4$ path integral as a character integral transform of a bulk Harish-Chandra character of the group $Spin(4,1)$ --  associated with the strictly massless fermionic discrete series UIRs with scaling dimension $\Delta=s+1$ and  spin $s$ \cite{Letsios:2023qzq, Anninos:2025mje, ottoson1968classification, schwarz1971unitary} -- and an edge character.  Note that in our previous work \cite{Anninos:2025mje}, the case of spin-$\tfrac{3}{2}$ was analysed in detail and a first conjecture for the character expression for the $S^4$ path integral of a generic spin-$s$ fermionic gauge field was presented. Our results verify this conjecture.
\newline\newline
Consider the one-loop $S^4$ path integral for a spin-$s \geq \tfrac{3}{2}$ totally symmetric fermionic gauge field,
\begin{equation} 
    \mathcal{Z} = \frac{\int \mathcal{D}\Psi~\mathcal{D}\Psi^\dagger \, e^{-S_E[\Psi]}}{\text{vol}(\mathcal{G})} ~.
\end{equation}
We conjecture that the path integral can be expressed as
\begin{equation} \label{assumption for generic spin strictly massless}
    \mathcal{Z}  = \Lambda_{\text{u.v.}}^{\mathcal{M}_s} \times \frac{\text{det}_{\text{TT}}^{(s)} \left( \frac{\slashed{\nabla}+i(s-\frac{1}{2})}{\Lambda_{\text{u.v.}}^{\frac{1}{2}}} \right)}{\text{det}_{\text{TT}}''^{\, (s-1)} \left( \frac{\slashed{\nabla}+i(s+\frac{1}{2})}{\Lambda_{\text{u.v.}}^{\frac{1}{2}}} \right)} ~,
\end{equation}
up to an overall numerical constant that is independent of $\Lambda_{\text{u.v.}}$, where $\mathcal{M}_s$ is a real number whose explicit value depends on the spin $s$. The double prime in $\text{det}_{\text{TT}}''^{\, (s-1)}$ denotes the omission of the two lowest-level eigenvalues of the operator $\slashed{\nabla}+i(s+\frac{1}{2})$, namely \cite{Homma:2020has, Letsios:2022tsq} $$\lambda_{\text{lowest}} = \pm i (s+\frac{1}{2}) +i(s+\frac{1}{2})~.$$ The conjecture (\ref{assumption for generic spin strictly massless}) has been verified for $s=\tfrac{5}{2}$ in section \ref{spin-5/2 section} and for $s = \tfrac{3}{2}$ in \cite{Anninos:2025mje}, and in both cases the explicit value of $\mathcal{M}_s$ has been computed explicitly via path integral methods. We will now proceed to express (\ref{assumption for generic spin strictly massless}) in terms of group-theoretic characters. We will also make a conjecture for the explicit value of $\mathcal{M}_s$ based on a preliminary analysis of the spin-$s \geq \frac{3}{2}$ path integral; as it turns out, such value ensures that the coefficient of the logarithmic UV divergence of the path integral is fully encoded by bulk and edge characters. A complete proof would require a path integral analysis akin to the one we give for the spin-$\tfrac{5}{2}$ case in section \ref{subsec gauge volume}, and for the spin-$\tfrac{3}{2}$ case in \cite{Anninos:2025mje}.
\newline\newline
We follow the technology developed in section 5 of \cite{Anninos:2020hfj}, and add some small variations throughout the presentation. Using standard heat kernel techniques, we express the logarithm of $\mathcal{Z}$ as
\begin{equation} 
    \log\mathcal{Z} = ~\log{\Lambda_{\text{u.v.}}^{\mathcal{M}_s}}- \int_{\mathbb{R}^+} \frac{d\tau}{\tau} \, e^{-\frac{\epsilon^2}{4\tau}} \left( \text{Tr}_{\text{TT}} \, e^{-\tau \left( \slashed{\nabla}_{s,\text{TT}} + m_s \right)} - \text{Tr}_{\text{TT}} \, e^{-\tau \left( \slashed{\nabla}_{s-1,\text{TT}} + \mu_{s-1} \right)} \right) ~,
\end{equation}
where $m_s=(s-\tfrac{1}{2})i$ and  $\mu_{s-1}=(s+\tfrac{1}{2})i$.\footnote{Spin-$(s-1)$ totally symmetric fermionic fields on dS$_4$ which satisfy the Dirac equation with imaginary mass parameter $\mu_{s-1}=(s+\tfrac{1}{2})i$ correspond to the so-called `shift symmetric' fermions \cite{Bonifacio:2023prb}.} The above can be re-expressed as
\begin{multline} \label{logZ}
    \log \mathcal{Z} =~\log{\Lambda_{\text{u.v.}}^{\mathcal{M}_s}} - \int_{\mathbb{R}^+} \frac{d\tau}{\tau} \, e^{-\frac{\varepsilon^2}{4\tau}} \underset{n\geq s-\frac{1}{2}}{\sum} \biggl( D^{5}_{n,s} \, e^{-\tau \left( (n+2)^2-(s-\frac{1}{2})^2 \right)} \\
    - D^{5}_{n,s-1} \, e^{-\tau \left( (n+2)^2-(s+\frac{1}{2})^2 \right)} \biggr) ~,
\end{multline}
where the dimension of the $Spin(5)$ representation furnished by spin-$s \in \{ \tfrac{1}{2}, \tfrac{3}{2}, \tfrac{5}{2}, \dots \}$ totally symmetric TT fermionic spherical harmonics on $S^4$ is given by  \cite{Homma:2020has}
\begin{equation}
    D^5_{n,s} = \frac{1}{3} (n+2)(2s+1)(n-s+\frac{3}{2})(n+s+\frac{5}{2}) ~.
\end{equation}
Recall that the Dirac operator acts on such TT fermionic spherical harmonics $\psi^{\pm}_{n|\mu_1 \dots \mu_s}$\footnote{These spherical harmonics carry $s-\frac{1}{2}$ totally symmetric tensor indices, and one spinor index which we suppress. They are gamma-traceless and divergence-free with respect to any of their indices.} on $S^4$ as \cite{Homma:2020has, Letsios:2022tsq, Letsios:2023qzq}
\begin{equation} \label{dirac spin s spectrum}
    \slashed{\nabla}\psi^{\pm}_{n|\mu_1 \dots \mu_{s-1/2}} = \pm i \, (n+2) \, \psi^{\pm}_{n|\mu_1 \dots \mu_{s-1/2}} ~, ~~~\text{with}~~~ n=s-\frac{1}{2}, s+\frac{1}{2},\dots ~,
\end{equation}
where the degeneracy of the eigenvalue $ \pm i \, (n+2)$ is given by $D^5_{n,s}$.
The allowed (or ``physical'') range of $n$ for a spin-$s$ spherical harmonic is $n \ge s-\tfrac{1}{2}$ \cite{Homma:2020has}. The ``unphysical'' range is $-1\leq n \leq s-\tfrac{3}{2}$, and it will shortly play an important role. Applying the Hubbard-Stratonovich trick, performing the $\tau$ integral and then a convenient contour rotation, as in \cite{Anninos:2020hfj, Anninos:2025mje}, we recast the sums appearing in the path integral as\footnote{Note that the physical range of the second sum is $n \geq s-\frac{1}{2}$ instead of $n \geq s-\frac{3}{2}$, since the lowest-level eigenvalues of the spin-$(s-1)$ determinant in (\ref{assumption for generic spin strictly massless}) are omitted.}
\begin{multline} \label{logZ after Hubd-Strat before extension}
    \log \mathcal{Z} = \log \Lambda_{\text{u.v.}}^{\mathcal{M}_s} 
     - \int_{\mathbb{R}^+}\frac{d t}{t}\Biggr(  \left( e^{-t(s-\frac{1}{2})} + e^{t(s-\frac{1}{2})} \right) \underset{n\geq s-\frac{1}{2}}{\sum} D^5_{n,s} \, e^{-t(n+2)} \\ - \left( e^{-t(s+\frac{1}{2})} + e^{t(s+\frac{1}{2})} \right) \underset{n\geq s-\frac{1}{2}}{\sum} D^5_{n,s-1} \, e^{-t(n+2)} \Biggr) ~,
\end{multline}
where we have formally set $\varepsilon = 0$. In order to derive the desired character formula, one needs to add and subtract the terms corresponding to the unphysical range in the sums. By doing so, we are able to re-express (\ref{logZ after Hubd-Strat before extension}) as
\begin{align} \label{logZ extended}
    \log \mathcal{Z} = \log \Lambda_{\text{u.v.}}^{\mathcal{M}_s} 
     & + \int_{\mathbb{R}^+}\frac{d t}{t}  \left( e^{-t(s-\frac{1}{2})} + e^{t(s-\frac{1}{2})} \right) \Bigg[-\underset{n\geq -1}{\sum} D^5_{n,s} \, e^{-t(n+2)}  + \sum^{s-\frac{3}{2}}_{n=-1} D^5_{n,s} \, e^{-t(n+2)} \Bigg] \nonumber \\ 
     & +\int_{\mathbb{R}^+}\frac{d t}{t} \left( e^{-t(s+\frac{1}{2})} + e^{t(s+\frac{1}{2})} \right) \left[ \underset{n\geq -1}{\sum} D^5_{n,s-1} \, e^{-t(n+2)}   - \sum^{s-\frac{3}{2}}_{n=-1} D^5_{n,s-1} \, e^{-t(n+2)}\right] ~.
\end{align}
For later convenience, note that the $n=s-\frac{5}{2}$ and the $n = s-\frac{3}{2}$ terms correspond, respectively, to zero modes of the spin-$s$ and spin-$(s-1)$ sectors, and that $D_{s-\frac{3}{2},s}^5 = 0 = D^5_{s-\frac{5}{2},s-1}$. At this point, we introduce the variable
\begin{align}
    q = e^{-t} ~,
\end{align}
and we recast the sums in (\ref{logZ extended}) as
\begin{align}\label{logZ very nice}
   \log \mathcal{Z}   = \log \frac{1}{\text{vol} \, \mathcal{G}_{\text{res}}}+\int_{\mathbb{R}^+} \frac{dt}{t} \left(\hat{F}_s(q) - F_{s}^0(q) \right) ~.
\end{align}
In the above,
\begin{equation} \label{F hat}
    \hat{F}_s(q) = - \left( q^{s-\frac{1}{2}} + q^{-s+\frac{1}{2}} \right) \underset{n\geq-1}{\sum} D^5_{n,s} \, q^{n+2} + \left( q^{s+\frac{1}{2}} + q^{-s-\frac{1}{2}} \right) \underset{n\geq-1}{\sum} D^5_{n,s-1} \, q^{n+2} ~,
\end{equation}
and $ {F}_{s}^0(q)$ encodes the subtraction of zero modes, namely
\begin{equation}
    {F}_{s}^0(q) =- D_{s-\frac{5}{2},s}^{5}\left(1+q^{2s-1}\right) + D_{s-\frac{3}{2},s-1}^{5}\left(1+q^{2s+1}\right) =  \frac{s(4s^2-1)}{3} \left( 2+q^{2s-1}+q^{2s+1} \right) ~.
\end{equation}
The coefficient of the logarithmic divergence associated with $-\int_{\mathbb{R}^+} \frac{dt}{t}  F_{s}^0(q) $ in (\ref{logZ very nice}) is found by taking the small-$t$ expansion: 
\begin{equation} \label{B coeff F0}
    \mathcal{B}_{F^0,s} = - \frac{4s(4s^2-1)}{3} ~.
\end{equation}
Removing zero modes by hand can result in a loss of locality. We avoid this issue by dividing by the locally defined volume of residual gauge transformations, $\text{vol} \, \mathcal{G}_{\text{res}}$.
These gauge transformations are generated by spin-$(s-1)$ Killing tensor-spinor gauge parameters, $K_{\mu_1 \dots \mu_{s-\frac{3}{2}}} = K_{(\mu_1 \dots \mu_{s-\frac{3}{2}})}$, which satisfy
\begin{align}
   \left( \nabla_{(\mu_1} + \frac{i}{2} \gamma_{(\mu_1} \right)K_{\mu_2 \dots \mu_{s-\frac{1}{2}})}=0 ~.
\end{align}
To be more specific, we defined
\begin{align} \label{vol G res integral formula}
    \log \frac{1}{\text{vol} \, \mathcal{G}_{\text{res}}}  \equiv  \log \Lambda_{\text{u.v.}}^{\mathcal{M}_s}  & + \int_{\mathbb{R}^+}\frac{d t}{t}  \left( q^{s-\frac{1}{2}} + q^{-s+\frac{1}{2}} \right) \left[ \sum^{s-\frac{7}{2}}_{n=-1} D^5_{n,s} \, q^{n+2} + D^5_{s-\frac{3}{2},s} \, q^{s+\frac{1}{2}} \right] \nonumber \\ 
    & -\int_{\mathbb{R}^+}\frac{d t}{t} \left( q^{s+\frac{1}{2}} + q^{-s-\frac{1}{2}} \right)     \sum^{s-\frac{5}{2}}_{n=-1} D^5_{n,s-1} \, q^{n+2}.
\end{align}
At this point, we are able to make an educated guess for the explicit value of $\mathcal{M}_s$. Let us first consider the left-hand side of the previous equation. Using our spin-$\frac{5}{2}$ analysis from section \ref{subsec gauge volume} and the spin-$\frac{3}{2}$ results of \cite{Anninos:2025mje}, one can conjecture that the $\Lambda_{\text{u.v.}}$ dependence of $\text{vol} \, \mathcal{G}_{\text{res}}$ for a generic spin-$s$ fermionic gauge field is given by
\begin{equation}\label{educated guess for volG_Res}
  \frac{1}{\text{vol} \, \mathcal{G}_{\text{res}}} \propto \Lambda_{\text{u.v.}}^{a+\frac{3}{2}a-\frac{1}{2}a} =  \Lambda_{\text{u.v.}}^{2a} ~,~~~ \text{where} ~~~ a = D^5_{s-\frac{3}{2},s-1} = \frac{s}{3}(4s^2-1) ~.
\end{equation}
In the above, the factor of $a$ comes from the different dimensionalities of the spin-$(s-1)$ fields which appear in the decomposition of the original spin-$s$ field, see for instance $\lambda_\mu^{(\text{TT})} \sim [\ell^{-\frac{1}{2}}]$ and $X_\mu^{(\text{TT})} \sim [\ell^{-\frac{3}{2}}]$ in (\ref{eq: residual volume 3/2}). The factor of $\tfrac{3}{2}a$ comes from the `volume' generated by half of the Killing tensor-spinor gauge parameters, i.e. the spin-$(s-1)$ modes such that $\slashed{\nabla}\epsilon_{\nu_1 ... \nu_{s-3/2}}=-i(s+\tfrac{1}{2})\epsilon_{\nu_1 ... \nu_{s-3/2}}$, as in (\ref{vol generated by KVS-}). Lastly, the factor of $-\tfrac{1}{2}a$ comes from the simplification of the ratio of two spin-$(s-1)$ functional determinants, appearing in $\mathcal{Z}_{\text{action}}$ and the Jacobian $J$, whereby the numerator excludes modes such that $\slashed{\nabla}\chi_{\nu_1 ... \nu_{s-3/2}}=i(s+\tfrac{1}{2})\chi_{\nu_1 ... \nu_{s-3/2}}$ and the denominator excludes modes such that $\slashed{\nabla}\chi_{\nu_1 ... \nu_{s-3/2}}=\pm i(s+\tfrac{1}{2})\chi_{\nu_1 ... \nu_{s-3/2}}$, see (\ref{Z action}) and (\ref{Z Jacobian}). Using (\ref{educated guess for volG_Res}), we find
\begin{equation} \label{B vol coeff}
    \mathcal{B}_{\text{vol},s} = 4a = \frac{4s(4s^2-1)}{3} ~,
\end{equation}
which is the coefficient of the logarithmic divergence associated with the residual gauge group volume. Now focus on the right hand side of (\ref{vol G res integral formula}). By taking the small-$t$ expansion of the integrand and keeping in mind that $\Lambda_{ \text{u.v.}} \propto \varepsilon^{-2}$, we find the same coefficient as in (\ref{B vol coeff}) for the logarithmic divergence provided that
\begin{equation} \label{value of M_s}
    \mathcal{M}_s = \frac{1}{96}(2s-1)(2s+1)(2s+3)(10s+1) ~.
\end{equation}
We conjecture that $\mathcal{M}_s$ is given by (\ref{value of M_s}) for $s \geq \frac{3}{2}$, and, as mentioned earlier, the cases $s=\frac{3}{2},\frac{5}{2}$ have been respectively verified in \cite{Anninos:2025mje} and the present paper. \\\\
Remarkably, we observe the cancellation
\begin{equation}
    \mathcal{B}_{\text{vol},s} +\mathcal{B}_{F^0,s} =0 ~,
\end{equation}
which entails that the coefficient of the logarithmic divergence of $\mathcal{Z}$ is purely encoded in $\int_{\mathbb{R}^+} \frac{dt}{t} \hat{F}_s(q)$. We will now express this last integral in terms of characters.

\paragraph{Naive characters and flipping.} So far we managed to express the path integral as
\begin{equation} \label{eq: non naive p.i.}
    \mathcal{Z} = \frac{1}{\text{vol} \, \mathcal{G}_{\text{res}}} \exp \left[  \int_{\mathbb{R}^+} \frac{dt}{t} \left( \hat{F}_{s}(q) - {F}_{s}^0(q) \right) \right] ~.
\end{equation}
After computing the sum over $n$ in (\ref{F hat}), we find 
\begin{multline}
    \hat{F}_s(q) = - \frac{q^{\frac{1}{2}}}{1-q} \biggl[ \left( 2(2s+1) \frac{q^{s+1}+q^{2-s}}{(1-q)^3} - \frac{2}{3}(s-\frac{1}{2})(s+\frac{1}{2})(s+\frac{3}{2}) \frac{q^s+q^{1-s}}{1-q} \right) \\ 
    - \left( 2(2s-1) \frac{q^{s+2}+q^{1-s}}{(1-q)^3} - \frac{2}{3}(s-\frac{3}{2})(s-\frac{1}{2})(s+\frac{1}{2}) \frac{q^{1+s}+q^{-s}}{1-q} \right) \biggr] ~.
\end{multline}
The above expression can be rewritten as
\begin{equation} \label{hatF}
    \hat{F}_s(q) = - \frac{q^{\frac{1}{2}}}{1-q} \left[ (\hat{\chi}_{\text{bulk},s} - \hat{\chi}_{\text{edge},s}) - (\hat{\chi}_{\text{bulk},s-1} - \hat{\chi}_{\text{edge},s-1}) \right] ~,
\end{equation}
where we defined the ``naive'' bulk and edge characters
\begin{equation}
    \hat{\chi}_{\text{bulk},s} \equiv 2(2s+1) \frac{q^{s+1}+q^{2-s}}{(1-q)^3} ~,
\end{equation}
\begin{equation}
    \hat{\chi}_{\text{edge},s} \equiv \frac{2}{3}(s-\frac{1}{2})(s+\frac{1}{2})(s+\frac{3}{2}) \frac{q^s+q^{1-s}}{1-q} ~,
\end{equation}
and
\begin{equation}
    \hat{\chi}_{\text{bulk},s-1} \equiv 2(2s-1) \frac{q^{s+2}+q^{1-s}}{(1-q)^3} ~,
\end{equation}
\begin{equation}
    \hat{\chi}_{\text{edge},s-1} \equiv \frac{2}{3}(s-\frac{3}{2})(s-\frac{1}{2})(s+\frac{1}{2}) \frac{q^{1+s}+q^{-s}}{1-q} ~.
\end{equation}
The integrand (\ref{hatF}) is clearly problematic, due to the appearance of negative powers of $q$ in the naive characters. We take care of this by implementing a fermionic analogue of the flipping procedure introduced in \cite{Anninos:2020hfj}. Given a naive bulk character, this procedure allows us to extract the true Harish-Chandra character corresponding to a unitary representation of $Spin(4,1)$. Similarly, given a naive edge character, this procedure allows us to extract the true edge character associated with representations of $Spin(2,1)$. In practice, the method involves flipping the signs of the coefficients and of the exponents of all the terms featuring a negative power of $q$ in the naive characters $\hat{\chi}$. \\\\
As an example, let us expand $\hat{\chi}_{\text{bulk},s}$ as
\begin{align}
    \hat{\chi}_{\text{bulk},s}  &= 2(2s+1) \sum_{m=0}^\infty \frac{(m+1)(m+2)}{2} (q^{m+s+1}+q^{m+2-s}) = \nonumber \\
    &=2(2s+1) \Biggr( \frac{q^{s+1}}{(1-q)^3}+\sum_{m=0}^\infty \frac{(m+1)(m+2)}{2} q^{m+2-s} \Biggr)~.
\end{align}
In the second line of the expression above, the terms with $m=0,\dots,s-\frac{5}{2}$ need to undergo the flipping procedure because the exponent $m+2-s$ becomes negative. After a bit of massaging, we are able to rewrite $\hat{\chi}_{\text{bulk},s}$ as
\begin{align}
     \hat{\chi}_{\text{bulk},s} =  [\hat{\chi}_{\text{bulk},s}] + 2(2s+1)   \sum_{m=0}^{s-\frac{5}{2}}\frac{(m+1)(m+2)}{2} (q^{s-2-m} +q^{-s+2+m}) ~,
\end{align}
where the corresponding flipped character is
\begin{multline}
    [\hat{\chi}_{\text{bulk},s}] = 2(2s+1) \sum_{m=0}^{s-\frac{5}{2}} \frac{(m+1)(m+2)}{2} (-q^{m+2-s}-q^{-m-2+s}) \\ + 2(2s+1) \sum_{m=0}^\infty \frac{(m+1)(m+2)}{2} (q^{m+s+1}+q^{m+2-s}) ~.
\end{multline}
Similarly,
\begin{align}
    \hat{\chi}_{\text{bulk},s-1}  
    =&2(2s-1) \Biggr( \frac{q^{s+2}}{(1-q)^3}+\sum_{m=0}^\infty \frac{(m+1)(m+2)}{2} q^{m+1-s} \Biggr)\\
    =&[ \hat{\chi}_{\text{bulk},s-1}] +2(2s-1) \sum_{m=0}^{s-3/2} \frac{(m+1)(m+2)}{2}( q^{s-1-m} + q^{-s+1+m})~,
\end{align}
where the flipped character is
\begin{align}
    [\hat{\chi}_{\text{bulk},s-1}] & = 2(2s-1) \sum_{m=0}^{s-\frac{3}{2}} \frac{(m+1)(m+2)}{2} (-q^{m+1-s}-q^{-m-1+s}) \nonumber \\ & + 2(2s-1) \sum_{m=0}^\infty \frac{(m+1)(m+2)}{2} (q^{m+s+2}+q^{m+1-s}) ~.
\end{align}
%
As far as the edge characters are concerned, we find similar expressions of the form
\begin{align}
&\hat{\chi}_{\text{edge},s}=[\hat{\chi}_{\text{edge},s}] + \dots~, \\
&  \hat{\chi}_{\text{edge},s-1}=[\hat{\chi}_{\text{edge},s-1}] + \dots  ~,
\end{align}
where the flipped characters are
\begin{multline}
    [\hat{\chi}_{\text{edge},s}] = \frac{2}{3}(s-\frac{1}{2})(s+\frac{1}{2})(s+\frac{3}{2}) \sum_{m=0}^{s-\frac{3}{2}} (-q^{m+1-s}-q^{-m-1+s}) \\ + \frac{2}{3}(s-\frac{1}{2})(s+\frac{1}{2})(s+\frac{3}{2}) \sum_{m=0}^\infty (q^{m+s}+q^{m+1-s}) ~,
\end{multline}
\begin{multline}
    [\hat{\chi}_{\text{edge},s-1}] = \frac{2}{3}(s-\frac{3}{2})(s-\frac{1}{2})(s+\frac{1}{2}) \sum_{m=0}^{s-\frac{1}{2}} (-q^{m-s}-q^{-m+s}) \\ + \frac{2}{3}(s-\frac{3}{2})(s-\frac{1}{2})(s+\frac{1}{2}) \sum_{m=0}^\infty (q^{m+s+1}+q^{m-s}) ~.
\end{multline}
We can now re-express (\ref{hatF}) as 
\begin{equation} \label{hat F towards the final line}
    \hat{F}_s(q) = - \frac{q^{\frac{1}{2}}}{1-q} \left (2{\chi}^{(s)}_{\text{bulk}} - 2{\chi}^{(s)}_{\text{edge}}\right) -\text{extra}  ~,
\end{equation}
%
where we identify the bulk Harish-Chandra character corresponding to the unitary fermionic discrete series representation of $Spin(4,1)$ labelled by scaling dimension $\Delta=s+1$ and spin $s$ \cite{hirai1965characters, Basile:2016aen, Anninos:2025mje}:
\begin{equation}\label{unitary bulk character flipped}
   2 \chi_{\text{bulk}}^{(s)} \equiv [\hat{\chi}_{\text{bulk},s}] - [\hat{\chi}_{\text{bulk},s-1}] = \frac{4}{(1-q)^3} \left( (2s+1)q^{s+1} - (2s-1)q^{s+2} \right) ~.
\end{equation}
We similarly identify the edge character
\begin{equation}\label{edge character flipped}
   2 \chi_{\text{edge}}^{(s)} \equiv [\hat{\chi}_{\text{edge},s}] - [\hat{\chi}_{\text{edge},s-1}] = \frac{4}{3(1-q)} (s-\frac{1}{2})(s+\frac{1}{2}) \left( (s+\frac{3}{2})q^s - (s-\frac{3}{2}) q^{s+1} \right) ~.
\end{equation}
The factors of $2$ in (\ref{unitary bulk character flipped}) and (\ref{edge character flipped}) have been introduced to emphasise that our fermions are complex (Dirac), leading to doubled degrees of freedom relative to the case of real fermions. We further note that $\chi_{\text{bulk}}^{(s)}$ can be expressed a sum of two characters corresponding to the direct sum of two discrete series UIRs of $Spin(4,1)$ of opposite helicity as
\begin{align}
   \chi_{\text{bulk}}^{(s)} = \chi_{\text{bulk}}^{(s),+}  + \chi_{\text{bulk}}^{(s),-} ~,
\end{align}
where
\begin{equation}
    \chi_{\text{bulk}}^{(s),+} = \frac{1}{(1-q)^3} \left( (2s+1)q^{s+1} - (2s-1)q^{s+2} \right) = \chi_{\text{bulk}}^{(s),-} ~.
\end{equation}
This is due to the fact that strictly massless fermions (as well as bosons) on dS$_4$ have two distinct helicities, and modes of fixed helicity separately furnish  discrete series UIRs of $Spin(4,1)$ \cite{Letsios:2023awz, Letsios:2022tsq, Letsios:2023qzq, Penedones:2023uqc, Higuchi:1991tn, RiosFukelman:2023mgq}.  
The bulk and edge characters in (\ref{unitary bulk character flipped}) and (\ref{edge character flipped}) agree with the prediction below formula (6.31) in \cite{Anninos:2025mje}, and with the results present in Appendix A of \cite{anninos2026ds4metamorphosis}. The term we call `extra'  in (\ref{hat F towards the final line}) is given by
\begin{align}
    \text{extra} = & - \frac{1}{12 (1 - q)^4} q^{\frac{1}{2} - s} \biggl[-3 + q^{3 + 2 s} (-3 + 2 s) (-1 + 2 s) (1 + 2 s) \nonumber \\
    & + q^3 (-1 + 2 s) (1 + 2 s) (3 + 2 s) + 2 s (1 + 6 s - 4 s^2) + 3 q (-1 + 2 s) (7 + 4 s^2) \\
    & - 3 q^{2 + 2 s} (-1 + 2 s) (7 + 4 s^2) - 3 q^2 (1 + 2 s) (7 + 4 s^2) + 3 q^{1 + 2 s} (1 + 2 s) (7 + 4 s^2) \nonumber \\
    & + q^{2 s} (3 - 2 s (-1 + 6 s + 4 s^2)) \nonumber \biggr] ~.
\end{align}
Note that the term `$\int \frac{dt}{t}\,\text{extra}$' does \textbf{not} contribute to the logarithmic divergence of the sphere path integral, as is evident from its small-$t$ expansion:
\begin{align}
    {\text{extra}}
     \approx -\frac{16~(4s^3-s)}{t} + O[t].
\end{align}

\paragraph{Final result.} It is convenient to single out the character contribution to the path integral by rewriting (\ref{eq: non naive p.i.}) as
\begin{align} \label{eq: non naive p.i. new}
    \mathcal{Z} &= \frac{1}{\text{vol} \, \mathcal{G}_{\text{res}}} \exp \left[  \int_{\mathbb{R}^+} \frac{dt}{t}  (\hat{F}_{s}(q) + \text{extra)}  \right] ~\exp \left[-  \int_{\mathbb{R}^+} \frac{dt}{t}   ({F}_{s}^0(q)+\text{extra})  \right] ~,\\
    &\equiv \frac{1}{{\text{vol} \, \mathcal{G}_{\text{res}}}}~  \mathcal{Z}_{\text{char},s}~\mathcal{Z}_{F^0,s} ~,
\end{align}
where
\begin{eqnarray}
   \log  \mathcal{Z}_{\text{char},s} &\equiv&    \int_{\mathbb{R}^+} \frac{dt}{t} \left( \hat{F}_s(q) + \text{extra} \right) ~,  \\
   \log  \mathcal{Z}_{F^0,s} &\equiv&  -\int_{\mathbb{R}^+} \frac{dt}{t} \left( {F}_s^0(q) + \text{extra} \right) ~.
\end{eqnarray}
Importantly, $ \mathcal{Z}_{\text{char},s}$ is expressed in terms of unitary bulk $Spin(4,1)$ characters (\ref{unitary bulk character flipped}) and edge characters (\ref{edge character flipped}), i.e.
\begin{equation} \label{final character formula}
    \log \mathcal{Z}_{\text{char},s} = - \int_{\mathbb{R}^+} \frac{dt}{t} \frac{e^{-\frac{t}{2}}}{1-e^{-t}} ~2\left( \chi_{\text{bulk}}^{(s)}(t) - \chi_{\text{edge}}^{(s)}(t) \right) ~.   
\end{equation}
As explained earlier, $\mathcal{Z}_{\text{char},s}$ fully encodes the coefficient of the logarithmic divergence of the one-loop $S^4$ path integral. By expanding $\mathcal{Z}_{\text{char},s}$ in the small-$t$ limit, the pre-coefficient of the $\tfrac{1}{t}$ divergent term is found to be
\begin{equation}\label{coef of log div any spin s fermion S^4}
    \mathcal{B}_{\text{final},s} =\mathcal{B}_{\text{char},s} = \left( \frac{17}{720} - \frac{5s^2}{6} +s^4 \right)_{\text{bulk},s} + \left( -\frac{1}{24} + \frac{2s^4}{3} \right)_{\text{edge},s} = -\frac{13}{720} - \frac{5s^2}{6} + \frac{5s^4}{3} ~.
\end{equation}
As a consistency check, letting $s=\tfrac{5}{2}$ we find 
\begin{equation}
    \mathcal{B}_{\text{char},\tfrac{5}{2}} = \left( \frac{3049}{90} \right)_{\text{bulk},\tfrac{5}{2}} + \left( 26 \right)_{\text{edge},\tfrac{5}{2}} = \frac{5389}{90} ~,
\end{equation}
which agrees with the result (\ref{eq: coeff log div func dets}) obtained from a direct path integral analysis -- without the use of the character formula  -- in the previous section. As a further consistency check, letting $s=\tfrac{3}{2}$ we find
\begin{equation}
    \mathcal{B}_{\text{char},\tfrac{3}{2}} = \left( \frac{289}{90} \right)_{\text{bulk},\tfrac{3}{2}} + \left( \frac{10}{3} \right)_{\text{edge},\tfrac{3}{2}} = \frac{589}{90} ~,
\end{equation}
in agreement with equation (6.30) of \cite{Anninos:2025mje}.

\paragraph{Coefficient of logarithmic divergence: dS vs AdS.} Consider the one-loop path integral for strictly massless complex fermionic gauge potentials of spin-$s \geq \tfrac{3}{2}$ on 4-dimensional Euclidean anti-de Sitter space, EAdS$_4$. A detailed study is missing from the literature, unlike the case of bosonic fields \cite{Sun:2020ame}. Nevertheless, bulk and edge character expressions for the Euclidean path integral of  strictly massless fermions on EAdS$_4$, like the ones we give in the present paper for $S^4$, are reported in equation (A.19) of \cite{anninos2026ds4metamorphosis}.\footnote{Further references which compute logarithmic divergence coefficients for strictly massless fermions on Euclidean AdS are \cite{Larsen:2015aia,Bobev:2023dwx}.} The coefficient of the logarithmic divergence for AdS$_4$ is equal to one half of the dS$_4$ coefficient (\ref{coef of log div any spin s fermion S^4}), namely
\begin{align}
    \mathcal{B}_{\text{final},s}^{\text{AdS}_4} = \frac{1}{2} \mathcal{B}_{\text{final},s} ~.
\end{align}
The reason is that  $\mathcal{B}_{\text{final},s}^{\text{AdS}_4}$ and $\mathcal{B}_{\text{final},s}$ are proportional to heat kernel coefficients and they are given by expressions of the form
\begin{align}
    \mathcal{B}_{\text{final},s}^{\text{AdS}_4} \propto \text{vol}(\text{EAdS}_4) ~~~\text{and}~~~ \mathcal{B}_{\text{final},s} \propto \text{vol}(S^4) ~,
\end{align}
where the proportionality constant is the same for both EAdS$_4$ and $S^4$, and the regulated volume of EAdS$_4$ is $\text{vol}(\text{EAdS}_4) = \tfrac{1}{2} \text{vol}(S^4)$.

\paragraph{Higher-spin spectra and bulk/edge cancellations.}

An interesting observation was made in \cite{Anninos:2025mje}, which later played a role in the gluing interpretation of the supersymmetric higher-spin sphere path integral in \cite{anninos2026ds4metamorphosis}. When one considers the $S^4$ path integral of an infinite tower of free strictly massless complex fermions of all spins $s=\tfrac{1}{2}, \tfrac{3}{2},\dots$, then the bulk and the edge characters cancel exactly.  Using our results (\ref{unitary bulk character flipped}), (\ref{edge character flipped}) and (\ref{final character formula}), we can readily verify this observation:%
\footnote{Note that the expressions for the bulk and edge characters, (\ref{unitary bulk character flipped}) and (\ref{edge character flipped}), also holds for $s=\tfrac{1}{2}$, corresponding to a massless Dirac spinor on $dS_4$.}%
\begin{equation} 
   \sum_{s=\tfrac{1}{2}}^{\infty} \log \mathcal{Z}_{\text{char},s} = - \int_{\mathbb{R}^+} \frac{dt}{t} \frac{e^{-\frac{t}{2}}}{1-e^{-t}} ~2~ \sum_{s=\tfrac{1}{2}}^{\infty}\left( \chi_{\text{bulk}}^{(s)}(t) - \chi_{\text{edge}}^{(s)}(t) \right) =0~.   
\end{equation}
Note that such a cancellation also happens for the same tower of fields placed on EAdS$_4$. This is easy to check using the results contained in Appendix A of \cite{anninos2026ds4metamorphosis}.



\section*{Acknowledgements}
It is a great pleasure to acknowledge Nicolas Boulanger, Beatrix Mühlmann, Stathis Vitouladitis, and especially Dionysios Anninos for many invaluable discussions. C.B. would like to thank the research unit `Physics of the Universe, Fields and Gravitation' of the University of Mons for hospitality, where part of this work was carried out. C.B. is funded by STFC under grant reference STFC/2887726. The work of V.A.L. is supported by the ULYSSE Incentive Grant for Mobility in Scientific Research [MISU] F.6003.24, F.R.S.-FNRS, Belgium.

\appendix

\section{Conventions and useful formulae} \label{appendix formulae}

Here we give some useful formulae concerning (tensor-)spinors on $S^4$ and dS$_4$.
The four gamma matrices $\gamma^\mu$ satisfy the anti-commutation relations
\begin{align}
    \gamma_\mu \gamma_\nu + \gamma_\nu \gamma_\mu= 2 g_{\mu \nu}\bm{1},
\end{align}
where $g_{\mu \nu}$ is the metric tensor\footnote{In Lorentzian signature, we use the mostly plus sign convention for the metric.} %
of $S^4$ or dS$_4$, depending on whether we are in Euclidean or Lorentzian signature, and $\bm{1}$ is the $4 \times 4$ spinorial identity matrix. 
We denote the flat 4-dimensional Minkowski metric as $\eta_{ab}=diag(-1,1,1,1)$, with $a,b \in \{ 0,1,2,3\}$, while the flat Euclidean metric is simply $\delta_{ab}$, with $a,b \in \{1,2,3,4\}$. We denote the anti-symmetrised products of gamma matrices as $\gamma^{\mu \nu} = - \gamma^{\nu \mu} =\gamma^{[\mu} \gamma^{\nu]}$ and  $\gamma^{\mu_1 \mu_2 \dots \mu_r} = \gamma^{[\mu_1}  \gamma^{\mu_2} \dots \gamma^{\mu_r ]}$ ($r=3,4$). Spinor indices are suppressed throughout this paper. \\\\
Gamma matrices $\gamma_\mu(x)$ carrying Greek vector indices are `curved-space gamma matrices'. They are related to the standard flat-space gamma matrices, which carry Latin vector indices, through the vierbein, as $\gamma_\mu(x) = e_{\mu}\,^b(x)~\gamma_b$. The vierbeins (and co-vierbeins) satisfy
\begin{align}
   & e_{\mu}\,^a e_{\nu}\,^b \eta_{ab}=g_{\mu \nu} ~, ~~~\text{on}~ \text{dS}_4 ~, \nonumber\\
    & e_{\mu}\,^a e_{\nu}\,^b \delta_{ab}=g_{\mu \nu} ~, ~~~\text{on}~ S^4 ~,
\end{align}
as well as $e^\mu\,_a~e_{\mu}\,^b = \delta_a\,^b$.  In Euclidean signature, we have $(\gamma^{b})^\dagger= \gamma^b$. In Lorentzian signature, we have $(\gamma^0)^\dagger= - \gamma^0$, while all spacelike gamma matrices are Hermitian. We denote the fifth gamma matrix with the same symbol, $\gamma^5$, in both Lorentzian and Euclidean signature. In either signature, this matrix is defined as the product of the other four gamma matrices (up to a phase). In our conventions, we have $(\gamma^5)^{\dagger}= \gamma^5$ and $(\gamma^{5})^2 = \bm{1}$ for both dS$_4$ and $S^4$. \\\\
Covariant derivatives $\nabla_\mu$ act on spinors $\psi$ as
\begin{equation}
    \nabla_\mu \psi = \partial_\mu \psi + \frac14\omega_\mu{}^{{ab}}\gamma_{{ab}} \, \psi ~,
\end{equation}
where $\omega_{\mu bc} = - \omega_{\mu cb}$ is the spin connection. The spin connection and the vierbein are related by
\begin{align}
  \partial_\mu e^{\rho}\,_b +\Gamma^\rho_{\mu \lambda} e^\lambda \,_b- \omega_{\mu}\,^c\,_b~e^\rho\,_{c}=0 ~,  
\end{align}
where $\Gamma_{ \mu \lambda}^\rho$ are the Christoffel symbols. Throughout our computations, we made use of the following useful identities:
\begin{itemize}
    \item $[\nabla^\alpha \nabla_\alpha, \slashed{\nabla}] \psi = 0 ~,$ 
    \item $[\nabla^\alpha \nabla_\alpha, {\nabla_\beta}] \psi = (\slashed{\nabla} \gamma_\beta + 2\nabla_\beta) \psi ~,$
    \item $\displaystyle [\slashed{\nabla}, \nabla_\beta] \psi = \frac{3}{2} \gamma_\beta \, \psi ~,$
    \item $\displaystyle [\nabla^\alpha , \slashed{\nabla}] \nabla_\alpha \psi = \frac{3}{2} \slashed{\nabla} \psi ~,$
    \item $\displaystyle [\slashed{\nabla}, \nabla_\alpha] \nabla_\beta \psi = ( \gamma_\beta \nabla_\alpha - g_{\alpha\beta} \slashed{\nabla} + \frac{3}{2} \gamma_\alpha \nabla_\beta) \psi ~,$
    \item $\displaystyle [\slashed{\nabla}^2, \nabla_\beta] \psi = 3 \nabla_\beta \, \psi ~.$ \\
\end{itemize}
The covariant derivative acts on vector-spinor fields $\psi_\mu$ as
\begin{equation}
    \nabla_\mu \psi_{\nu} = \partial_\mu \psi_{\nu} + \frac{1}{4} \omega_\mu{}^{{ab}} \gamma_{{ab}} \psi_{\nu}  - \Gamma^\lambda_{\mu\nu}\psi_{\lambda} ~,
\end{equation}
and the generalisation to the case of tensor-spinors with more tensor indices is straightforward.
The commutator of covariant derivatives acting on a spinor $\psi$ on the unit $S^4$ (or dS$_4$) is given by
\begin{equation}\label{commutator of cov derivs spinor}
    \left[ \nabla_\mu, \nabla_\nu \right] \psi = \frac{1}{2} \gamma_\mu \gamma_\nu \psi - \frac{1}{2} g_{\mu\nu} \psi ~.
\end{equation}
Moreover, for the spinor $\psi$ the following relation holds:
\begin{equation} 
    \nabla^\alpha \nabla_\alpha \psi = \slashed{\nabla}^2 \psi + \frac{R}{4} \psi = \slashed{\nabla}^2 \psi + 3\psi ~,
\end{equation}
where $R=12$ is the scalar curvature of the unit $S^4$ (or dS$_4$). The commutator of covariant derivatives acting on a vector-spinor $\psi_\mu$ on the unit $S^4$ (or dS$_4$) is given by
\begin{equation}\label{commutator of cov derivs vec-spinor}
    \left[ \nabla_\mu, \nabla_\nu \right] \psi_\rho = \frac{1}{2} \gamma_\mu \gamma_\nu \psi_\rho - \frac{1}{2} g_{\mu\nu} \psi_\rho + 2g_{\rho [\mu} \psi_{\nu]} ~.
\end{equation}
Moreover, for gamma-traceless vector-spinors $\psi_\mu$, we have
\begin{equation} 
    \nabla^\alpha \nabla_\alpha \psi_{{\mu}} = \slashed{\nabla}^2 \psi_{{\mu}} + \left( \frac{R}{4} +1 \right)\psi_{{\mu}} = \slashed{\nabla}^2 \psi_{{\mu}} + 4\psi_{{\mu}} ~.
\end{equation}
The commutator of covariant derivatives acting on a tensor-spinor $\psi_{\mu\nu}$ on the unit $S^4$ (or dS$_4$) is given by
\begin{equation}
    \left[ \nabla_\mu, \nabla_\nu \right] \psi_{\rho \sigma} = \frac{1}{2} \gamma_\mu \gamma_\nu \psi_{\rho\sigma} - \frac{1}{2} g_{\mu\nu} \psi_{\rho\sigma} + 2g_{\rho [\mu} \psi_{\nu]\sigma} + 2g_{\sigma [\mu} \psi_{|\rho| \nu]} ~.
\end{equation}

\section{Conformal Killing vector-spinors on $S^4$} \label{App: conformal Killing vec-spin}

We define conformal Killing vector-spinors (CKVSs) on the unit $S^4$ to be vector-spinors $C^{\pm}_{\mu}$ which satisfy
\begin{equation} \label{def: CKV-spinor equation}
    \left( \nabla_{(\mu}  \pm \frac{i}{2}  \gamma_{(\mu} \right) C^{\mp}_{\nu)} = \frac{g_{\mu \nu}}{4}  \left(\nabla^{\rho}  \pm \frac{i}{2}  \gamma^{\rho} \right) C^{\mp}_{\rho} ~.
\end{equation}
The two families of conformal Killing vector-spinors are related to each other by
\begin{equation}
    C^{\pm}_{\mu} = \gamma^5 C^{\mp}_{\mu} ~.
\end{equation}
We can further distinguish two categories of CKVSs:
\begin{itemize}
    \item \textit{(isometric) Killing vector-spinors} -- these are transverse and gamma-traceless conformal Killing vector-spinors, $C^{\pm}_{\mu} = A^{\pm}_{\mu}$, which satisfy $$\left( \nabla_{(\mu}  \pm \frac{i}{2}  \gamma_{(\mu} \right) A^{\mp}_{\nu)} = 0 ~;$$
    
    \item \textit{non-isometric conformal Killing vector-spinors} -- these are conformal Killing vector-spinors, $C^{\pm}_{\mu} = V^{\pm}_{\mu}$, such that $\gamma^{\mu}V^{\pm}_{\mu} \neq 0$ and $\nabla^{\mu}V^{\pm}_{\mu} \neq 0$.
\end{itemize}
Interestingly, non-isometric CKVSs can always be expressed in terms of certain spinors $\phi^{\pm}$, which we call ``spinor potentials'', as 
\begin{equation} \label{non isometric CKVSs}
    V^{\mp}_{\mu} = \left(\nabla_{\mu} \pm \frac{i}{2} \gamma_{\mu} \right) \phi^{\mp} ~.
\end{equation}
This is similar to the fact that non-isometric conformal Killing vectors on spheres can be expressed as gradients of certain scalar potentials -- see e.g. \cite{Letsios:2023awz}.
Note that the spinors $\phi^{\pm}$ are not conformal Killing spinors. 
However, these spinors are not arbitrary: they belong to a certain class of spinorial spherical harmonics on $S^4$. In particular, they satisfy
\begin{equation}
    \slashed{\nabla} \phi^{\pm} = \pm i(n+2) \phi^{\pm} ~,~~~\text{with} ~~~ n=1 ~.
\end{equation}
These are once again related to each other by $\gamma^{5}\phi^{\pm} = \phi^{\mp}$, and they satisfy
\begin{equation}
    \left( \nabla_{(\mu}\nabla_{\nu)} \pm i \gamma_{(\mu} \nabla_{\nu)} + \frac{3}{4}g_{\mu \nu}  \right) \phi^{\mp}=0 ~.
\end{equation}
Note that $\nabla^2 \phi^{\pm} = -6 \phi^{\pm}$. In fact, on $S^4$
\begin{equation} \label{spinors forming CKVSs}
    \left( \nabla_{(\mu}\nabla_{\nu)} \pm i \gamma_{(\mu} \nabla_{\nu)} + \frac{3}{4}g_{\mu \nu}  \right) \phi^{\mp}=0 ~~~ \Leftrightarrow ~~~ \slashed{\nabla} \phi^{\pm} = \pm 3i \phi^{\pm} ~.
\end{equation}
\paragraph{Proof of equation (\ref{spinors forming CKVSs}).}
Let us first prove that the following holds:
\begin{equation} 
    \left( \nabla_{(\mu}\nabla_{\nu)} \pm i \gamma_{(\mu} \nabla_{\nu)} + \frac{3}{4}g_{\mu \nu}  \right) \phi^{\mp}=0 ~~~ \Leftarrow ~~~ \slashed{\nabla} \phi^{\pm} = \pm 3i \phi^{\pm} ~.
\end{equation}
We define the symmetric tensor-spinors $O^+_{\mu \nu}$ and $O^{-}_{\mu \nu}$ as
\begin{equation} 
  O^{\mp}_{\mu \nu} \equiv  \left( \nabla_{(\mu}\nabla_{\nu)} \pm i \gamma_{(\mu} \nabla_{\nu)} + \frac{3}{4}g_{\mu \nu}  \right) \phi^{\mp}  ~,
\end{equation}
where  $\phi^{\mp}$ satisfies $\slashed{\nabla} \phi^{\pm} = \pm 3i \phi^{\pm}$. One can show that
\begin{align}
    \int_{S^4} d^4x~\sqrt{g}~O^{+~\dagger}_{\mu \nu}~O^{+\mu \nu} = 0 = \int_{S^4} d^4x~\sqrt{g}~O^{-~\dagger}_{\mu \nu}~O^{-\mu \nu} ~,
\end{align}
by expressing $O^{\pm}_{\mu \nu}$ in terms of $\phi^{\pm}$ and performing some integrations by parts. Since the inner product on $S^4$ is positive definite, we conclude that $O^{\pm}_{\mu \nu}=0$ as a consequence of its vanishing norm. Similarly, it is easy to prove that
\begin{equation} 
    \left( \nabla_{(\mu}\nabla_{\nu)} \pm i \gamma_{(\mu} \nabla_{\nu)} + \frac{3}{4}g_{\mu \nu}  \right) \phi^{\mp}=0 ~~~ \Rightarrow ~~~ \slashed{\nabla} \phi^{\pm} = \pm 3i \phi^{\pm} ~.
\end{equation}
To be specific, if one takes the trace of the left-hand side, as
\begin{equation} 
   g^{\mu\nu} \left( \nabla_{(\mu}\nabla_{\nu)} \pm i \gamma_{(\mu} \nabla_{\nu)} + \frac{3}{4}g_{\mu \nu}  \right) \phi^{\mp}=(\slashed{\nabla}^2  \pm i \slashed{\nabla} + 6)\phi^{\mp}=0 ~,
\end{equation}
it is easy to see that $\phi^{\mp}$ is an eigenfunction of the Dirac operator $\slashed{\nabla}$ with eigenvalue $\mp 3i $. 
\textbf{End of proof.}%
\newline\newline
Defining non-isometric CKVSs as in (\ref{non isometric CKVSs}), we find that equation (\ref{spinors forming CKVSs}) is equivalent to the conformal Killing vector-spinor equation (\ref{def: CKV-spinor equation}). Moreover, a straightforward calculation shows that
\begin{equation}
   \left(\nabla_{\nu} \pm \frac{i}{2} \gamma_{\nu} \right)  V^{\mp}_{\mu} =  \left(\nabla_{\mu} \pm \frac{i}{2} \gamma_{\mu} \right)  V^{\mp}_{\nu} ~.
\end{equation}
Thus, the CKVSs equation (\ref{def: CKV-spinor equation}) can be re-expressed as 
\begin{equation}
    \left(\nabla_{\mu}  \pm \frac{i}{2}  \gamma_{\mu} \right) V^{\mp}_{\nu} = \frac{g_{\mu \nu}}{4}  \left(\nabla^{\rho}  \pm \frac{i}{2}  \gamma^{\rho} \right)V^{\mp}_{\rho} ~,
\end{equation}
and one can easily verify that
\begin{equation}
    \left (\nabla^{\rho} \pm \frac{i}{2}  \gamma^{\rho} \right)V^{\mp}_{\rho} = -4 \phi^{\mp} ~.
\end{equation}
\newline
\textbf{Justifying the name ``(isometric) Killing vector-spinors''.} The name stems from the fact that `isometric' Killing vector-spinors (KVSs) can be used to construct Killing tensors of rank 2. For instance, if one considers two KVSs $A^{(1)}_\mu$ and $A^{(2)}_\nu$ such that $$\left( \nabla_{(\mu} + \frac{i}{2} \gamma_{(\mu} \right)A^{(1)}_{\nu)} = 0 = \left( \nabla_{(\mu} + \frac{i}{2} \gamma_{(\mu} \right)A^{(2)}_{\nu)} ~,$$ then it is possible to show that the symmetric rank-2 tensor $K_{\mu\nu}$ defined as
\begin{equation}
    K_{\mu\nu} \equiv A^{(1)\dagger}_{(\mu} A^{(2)}_{\nu)} ~,
\end{equation}
(where the spinor indices of $A^{(1)\dagger}_{\mu}$ and $A^{(2)}_{\nu}$ are contracted) satisfies the Killing tensor equation, i.e. $$\nabla_{(\alpha} K_{\mu\nu)} = 0 ~.$$ Indeed, after a little index gymnastics, one finds that
\begin{align}
    \nabla_{(\alpha} K_{\mu\nu)} & = \left( \nabla_{(\alpha}A^{(1)\dagger}_\mu \right) A^{(2)}_{\nu)} + A^{(1)\dagger}_{(\alpha} \nabla_\mu A^{(2)}_{\nu)} = \nonumber \\
    & = \frac{i}{2} A^{(1)\dagger}_{(\alpha} \gamma_\mu A^{(2)}_{\nu)} + A^{(1)\dagger}_{(\alpha} \left( - \frac{i}{2} \right) \gamma_\mu A^{(2)}_{\nu)} = 0 ~,
\end{align}
where in the second line we used the fact that $A^{(1)}_\mu$ and $A^{(2)}_\nu$ are KVSs.
\newline\newline
\textbf{Justifying the name ``non-isometric conformal Killing vector-spinors''.} Analogously to the previous case, the name stems from the fact that non-isometric CKVSs can be used to construct conformal Killing tensors of rank 2 that are not Killing tensors. In other words, non-isometric CKVSs can be used to construct what we may call ``non-isometric conformal Killing tensors'' of rank 2. In fact, if one considers two CKVSs $V^{(1)}_\mu$ and $V^{(2)}_\nu$ which satisfy (\ref{def: CKV-spinor equation}), it is possible to show that the totally symmetric rank-2 tensor $C_{\mu\nu}$ defined as
\begin{equation}\label{construct non-isom conf Kill tens for non-isom conf Kill vec-spin}
    C_{\mu\nu} \equiv V^{(1)\dagger}_{(\mu} V^{(2)}_{\nu)} ~,
\end{equation}
satisfies the conformal Killing tensor equation, i.e.%
$$\nabla_{(\alpha} C_{\mu\nu)} = \frac{1}{D+2} g_{(\alpha\mu} \left( 2\nabla^\beta C_{\beta\nu)} + \nabla_{\nu)} C^\beta_\beta \right) ~,$$%
with $D=4$ in our case. Indeed, after a lengthy calculation, one finds that
\begin{align}
    \nabla_{(\alpha} C_{\mu\nu)} & = \frac{1}{4} g_{(\alpha\mu} \left[ \left( \nabla \cdot V^{(1)\dagger} - \frac{i}{2} V^{(1)\dagger} \cdot \gamma \right) V^{(2)}_{\nu)} + V^{(1)\dagger}_{\nu)} \left( \nabla \cdot V^{(2)} + \frac{i}{2} \gamma \cdot V^{(2)} \right) \right] = \nonumber \\
    & = \frac{1}{6} g_{(\alpha\mu} \left( 2\nabla^\beta C_{\beta\nu)} + \nabla_{\nu)} C^\beta_\beta \right) ~,
\end{align}
where we used the fact that $V^{(1)}_\mu$ and $V^{(2)}_\nu$ are CKVSs. Thus, $C_{\mu \nu}$ satisfies the conformal Killing tensor equation.


\section{Details on zero modes excluded from path integration} \label{Appendix: excluded modes}

Here we expand further on the discussion below (\ref{path integral 1/2 from action}), in regard to the zero modes that we exclude from path integration. In order to derive equation (\ref{sphere path integral 5/2 new}) from the initial path integral (\ref{sphere path integral 5/2}), one must decompose the field as in (\ref{vasilis pretty equation}), introduce an appropriate Jacobian factor $J$ and then temporarily exclude zero modes so as to render path integration well-defined. Such zero modes, however, must still be path integrated over separately.%
\footnote{This is in contrast with modes excluded from path integration due to the requirement that the field decomposition (\ref{vasilis pretty equation}) is not redundant. We do not integrate over those modes and locality is preserved.} 
More specifically, in (\ref{sphere path integral 5/2 new}) the integration measures have been split into their zero mode contributions and the remainder, as
\begin{equation}
    \int \mathcal{D} {X}^{(\text{TT})} \mathcal{D} {X}^{(\text{TT})\dagger}= \int   \mathcal{D}' {X}^{(\text{TT})} \mathcal{D}' {X}^{(\text{TT})\dagger}\times \int_{ker(\slashed{\nabla}-3i)} \mathcal{D} {X}^{(\text{TT})}  \mathcal{D} {X}^{(\text{TT})\dagger} ~,
\end{equation}
\begin{equation}
    \int \bar{\mathcal{D}} P \bar{\mathcal{D}} P^\dagger= \int \bar{\mathcal{D}}^{*} P \bar{\mathcal{D}}^{*} P^\dagger \times \int_{ker(\slashed{\nabla}-6i)} \mathcal{D} P~\mathcal{D} P^\dagger ~,
\end{equation}
\begin{equation}
    \int \tilde{\mathcal{D}} B \tilde{\mathcal{D}} B^\dagger= \int \tilde{\mathcal{D}}^{\star} B~\tilde{\mathcal{D}}^{\star} B^\dagger  \times \int_{ker(\slashed{\nabla}+6i)} \mathcal{D} B~\mathcal{D} B^{\dagger} ~.
\end{equation}
Throughout the paper, we equivalently use the notation\footnote{Remember that $P^*$ is not the complex conjugate of $P$, and $B^\star$ is not the complex conjugate of $B$.}
\begin{equation}
    \int_{ker(\slashed{\nabla}-3i)} \mathcal{D} {X}^{(\text{TT})}  \mathcal{D} {X}^{(\text{TT})\dagger} \equiv   \int\mathcal{D} X'^{\,(\text{TT})}\mathcal{D} X'^{\,(\text{TT})\dagger} ~,
\end{equation}
\begin{equation}
    \int_{ker(\slashed{\nabla}-6i)} \mathcal{D} P \mathcal{D} P^\dagger \equiv \int \mathcal{D} P^* \mathcal{D} P^{*\dagger} ~,
\end{equation}
\begin{equation}
    \int_{ker(\slashed{\nabla}+6i)} \mathcal{D} B~\mathcal{D} B^{\dagger} \equiv   \int \mathcal{D} B^\star \mathcal{D} B^{\star \dagger} ~.
\end{equation}
In order to further clarify the notation, in (\ref{path integral 1/2 from action}) the determinant 
\begin{align}\label{explaning det in appendix}
 \widetilde{\overline{\text{det}}}^{* \star \, (\frac{1}{2})} \, \Big({(\slashed{\nabla}^2+4)(\slashed{\nabla}^2+9)} \Big)  
\end{align}
denotes the determinant of the corresponding differential operator acting on spinors, whereby the eigenvalues associated with the symbols $^*$, $^\star$, $-$ and $\sim$ have been omitted. In other words, we omitted the spin-$\tfrac{1}{2}$ modes in $ker(\slashed{\nabla}-6i)$, $ker(\slashed{\nabla}+6i)$, $ker(\slashed{\nabla}+2i) \cup ker(\slashed{\nabla}-2i)$ (Killing spinors, see (\ref{def: Killing spinors very nice})) and $ker(\slashed{\nabla}+3i) \cup ker(\slashed{\nabla}-3i)$ (spinor potentials, see (\ref{def: spinor potentials very nice})) respectively. \\\\
We also use the same symbols, $^*$, $^\star$, $-$ and $\sim$, to label a spinor, say $\psi$, which belongs to the aforementioned kernels:
\begin{align}
    & \psi^* \in ker(\slashed{\nabla}-6i) ~, \nonumber \\ 
    & \psi^{\star} \in ker(\slashed{\nabla}+6i) ~, \nonumber\\
    & \bar{\psi}  \in ker(\slashed{\nabla}+2i) \cup ker(\slashed{\nabla}-2i) ~,~~~ \text{i.e.}~\bar{\psi}~\text{is a Killing spinor (\ref{def: Killing spinors very nice})} ~, \nonumber\\
    & \tilde{\psi} \in ker(\slashed{\nabla}+3i) \cup ker(\slashed{\nabla}-3i) ~,~~~ \text{i.e.}~\tilde{\psi}~\text{is a spinor potential (\ref{def: spinor potentials very nice})} ~.
\end{align}
Similarly, we combine different labels on $\psi$ to denote that $\psi$ belongs to the union of the corresponding kernels. For example:
\begin{align} \label{clarifying kernels}
  & \tilde{\psi}^\star \in ker(\slashed{\nabla}+3i) \cup ker(\slashed{\nabla}-3i) \cup ker(\slashed{\nabla}+6i) \nonumber ~, \\
  & \bar{\psi}^* \in ker(\slashed{\nabla}+2i) \cup ker(\slashed{\nabla}-2i) \cup ker(\slashed{\nabla}-6i) ~.
\end{align}
Following these notational conventions, in  the determinant (\ref{explaning det in appendix}) (see also (\ref{path integral 1/2 from action})) the excluded modes are $\bar{P}, P^*, P^\star, \tilde{P} $ and  $\bar{B}, B^*, B^\star, \tilde{B} $. As mentioned in the main text, the exclusion of the modes $\bar{P}$ and $\tilde{B}$ stems from the requirement that the field decomposition (\ref{vasilis pretty equation}) has no redundancies (or in other words, these modes must not be path integrated over). The modes $P^{*}$ and $B^{\star}$ are instead momentarily excluded because they are zero modes of (\ref{path integral 1/2 from action}), but they  are still path integrated over in $\mathcal{Z}_{\text{gauge vol}}$, see (\ref{define Z_gauge vol}). In order to arrive at the expression in the last line of (\ref{path integral 1/2 from action}), which includes the determinant (\ref{explaning det in appendix}), we need to have the same range of path integration for both $P$ and $B$. In order to achieve this, we also exclude $P^\star$, $\tilde{P}$, $B^*$ and $\bar{B}$ from the functional determinant computation, and we carry out their path integration separately in (\ref{Chiara's compromise for f(B,rho etc)}). \\\\
To carry out the computation of the function $f(\bar{B},B^*,\tilde{P},P^{\star})$ in (\ref{Chiara's compromise for f(B,rho etc)}), one can expand the spinor fields $B$ and $P$ in spherical harmonics $\psi_n^{\pm}$ on $S^4$ as\footnote{To be precise, the mode expansions should be written as
\begin{equation}
    B = \sum_{n=0}^{\infty} \sum_{k=1}^{D_{n,\frac{1}{2}}^5} (c_{n,k}^{+} \, \psi_{n,k}^{+} + c_{n,k}^{-} \, \psi_{n,k}^{-})~, ~~~\text{and}~~~ P = \sum_{n=0}^{\infty} \sum_{k=1}^{D_{n, \frac{1}{2}}^5} (d_{n,k}^{+} \, \psi_n^{+} + d_{n,k}^{-} \, \psi_n^{-}) ~,
\end{equation}
where the label $k=1, \dots ,D^5_{n,\frac{1}{2}}$ keeps track of the degeneracy of each eigenspace with fixed $n$ \cite{Camporesi:1995fb}. We often omit the label $k$ for notational simplicity.}
\begin{equation}
    B = \sum_{n=0}^{\infty} c_n^{\pm} \, \psi_n^{\pm} ~, ~~~\text{and}~~~ P = \sum_{n=0}^{\infty} d_n^{\pm} \, \psi_n^{\pm} ~.
\end{equation}
Then, as clarified by (\ref{clarifying kernels}) and keeping (\ref{dirac spin 1/2 spectrum}) in mind, the $\bar{B}$ modes correspond to the spherical harmonics $\psi_0^{\pm}$, the $B^*$ modes correspond to the spherical harmonics $\psi_4^{+}$, the $\tilde{P}$ modes correspond to the spherical harmonics $\psi_1^{\pm}$   and the ${P}^\star$ modes correspond to the spherical harmonics $\psi_4^{-}$. If we take a closer look at the structure of $f(\bar{B},B^*,\tilde{P},P^{\star})$ and examine the $B$-dependent part more in detail, we can rewrite it as
\begin{multline}
    \int \mathcal{D} \bar{B} \mathcal{D} {\bar{B}}^{\dagger} \mathcal{D} B^{*} \mathcal{D} B^{* \, \dagger} \exp{\Bigg( 4 \int d^4x \sqrt{g}~ B^\dagger (\slashed{\nabla} + 6i) B\Bigg)} = \\
    \int d c_{0}^+ d c_{0}^{+ \,*} d c_{0}^- d c_{0}^{- \,*}  ~\exp{\Bigg( 4\times 8i \times c_{0}^+ \, {c_{0}^+}^{*} \Bigg)} \times \exp{\Bigg( 4\times 4i \times c_{0}^-  {c_{0}^-}^{*} \Bigg)} \times \\
    \int d c_{4}^+ d c_{4}^{+ \,*}  ~\exp{\Bigg( 4\times 12i \times c_{4}^+ \, {c_{4}^+}^{*} \Bigg)} ~.
\end{multline}
One can similarly expand the $P$-dependent part of $f(\bar{B},B^*,\tilde{P},P^{\star})$, and it is then straightforward to obtain (\ref{extra modes from action VASILIS}).
%

\section{Symmetric tensor-spinor spherical harmonics on $S^4$} \label{appendix: longitudinal modes}

In this Appendix, we show that the decomposition of the symmetric tensor-spinor $\Psi_{\mu \nu}$ on $S^4$ given in (\ref{vasilis pretty equation}) follows from the completeness of tensor-spinor spherical harmonics. It is convenient to begin our analysis from the decomposition (\ref{psi decomposition 2 generic}), which we write here for convenience
%
\begin{align} \label{decomposition appendix initial  TT+long}
    \Psi_{\mu\nu} = \Psi_{\mu\nu}^{(\text{TT})} + \Psi^{(\text{long})}_{\mu \nu} ~.    
\end{align}
The transverse-traceless (TT) part $\Psi_{\mu\nu}^{(\text{TT})}$ is divergence-free and gamma-traceless with respect any of its tensor indices, while the longitudinal term can be further expanded as
\begin{align}  \label{decomposition appendix initial long only}
    \Psi^{(\text{long})}_{\mu \nu} =\Psi^{(\text{long}| \text{spinor})}_{\mu \nu} + \Psi^{(\text{long}|\text{vec-spinor})}_{\mu \nu} ~.
\end{align}
The longitudinal component $\Psi^{(\text{long}| \text{vec-spinor})}_{\mu \nu}$ is built out of TT vector-spinors on $S^4$, as
\begin{align}\label{decomposition appendix longitudinal vec-spinor}
    \Psi^{(\text{long}| \text{vec-spinor})}_{\mu \nu} = \nabla_{(\mu} \lambda^{(\text{TT})}_{\nu)} + \gamma_{(\mu} U^{(\text{TT})}_{\nu)} ~,
\end{align}
where $\lambda^{(\text{TT})}_{\nu}$ is orthogonal to all Killing vector-spinors (\ref{def: Killing vec-spinors very nice}) on $S^4$, while 
$U^{(\text{TT})}_{\nu}$ is a generic TT vector-spinor. 
The longitudinal component $\Psi^{(\text{long}| \text{spinor})}_{\mu \nu}$ is instead expressed in terms of spinors on $S^4$, as
\begin{align}\label{decomposition appendix longitudinal spinor}
    \Psi^{(\text{long}| \text{spinor})}_{\mu \nu} = \nabla_{(\mu} \nabla_{\nu)} \xi + \gamma_{(\mu} \nabla_{\nu)} M +g_{\mu \nu}\Lambda~,
\end{align}
where each of the three spinors  $\xi, M, \Lambda$ is orthogonal to all Killing spinors (\ref{def: Killing spinors very nice}) on $S^4$. Moreover, the spinor $\Lambda$ is orthogonal to all spinor potentials (\ref{def: spinor potentials very nice}).

\subsection{TT and longitudinal spherical harmonics} On $S^4$, there exist two distinct types of symmetric tensor-spinor spherical harmonics: the transverse-traceless (TT) and the longitudinal ones. Both of these are eigenfunctions of the Dirac operator $\slashed{\nabla}$ on $S^4$, as we explain below. The TT tensor-spinor harmonics have been constructed explicitly in \cite{Letsios:2022tsq}, while, to the best of our knowledge, the longitudinal harmonics will be constructed in the present Appendix for the first time. 

\paragraph{TT harmonics.} Let us denote the TT symmetric tensor-spinor spherical harmonics as $\psi^{\pm}_{n|\mu\nu}$, where we suppress some extra labels for convenience (let $s=\frac{5}{2}$ in (\ref{dirac spin s spectrum})).
These tensor-spinors satisfy
\begin{align}
    \slashed{\nabla}\psi^{\pm}_{n|\mu\nu} = \pm i (n+2) \psi^{\pm}_{n|\mu\nu} ~,~~~\text{and}~~~ \gamma^\mu \psi^{\pm}_{n|\mu\nu} = 0 = \nabla^\mu \psi^{\pm}_{n|\mu\nu} ~,
\end{align}
where the quantum number $n$ takes the values $n \in \{2,3,...\}$.
For a detailed discussion of TT spherical harmonics see \cite{Letsios:2022tsq}.

\paragraph{Longitudinal harmonics.} As for the longitudinal symmetric tensor-spinor spherical harmonics, we will denote them as  $v^{(1)\pm}_{m|\mu\nu}, ~v^{(2)\pm}_{m|\mu\nu}$ and $\lambda^{(1)\pm}_{m|\mu\nu}$, $\lambda^{(2)\pm}_{m|\mu\nu}$ (the notation will be further clarified shortly). The quantum number $m \in \mathbb{N}$ labels the eigenvalue of the Dirac operator and its allowed range will be given explicitly below. We have suppressed extra labels for convenience.
 
\paragraph{Completeness.} The set of TT and longitudinal harmonics forms a complete set of modes for symmetric tensor-spinors on $S^4$. We will exploit this to show that a generic complex tensor-spinor $\Psi_{\mu \nu}$ can be decomposed as in (\ref{decomposition appendix initial  TT+long})-(\ref{decomposition appendix longitudinal spinor}). Due to the completeness of the spherical harmonics, $\Psi_{\mu \nu}$ can be always expanded as
\begin{align}
    \Psi_{\mu \nu} =\sum_{n=2}^{\infty}\sum_{\sigma = \pm }a^{\sigma}_n \, \psi^{\sigma}_{n|\mu\nu} + \sum_{m}\sum_{\sigma = \pm }\sum_{j=1,2}b^{(j)\sigma}_m \, v^{(j)\sigma}_{m|\mu\nu}+ \sum_{m}\sum_{\sigma = \pm }\sum_{j=1,2}c^{(j)\sigma}_m \, \lambda^{(j)\sigma}_{m|\mu\nu}~,
\end{align}
 where the sums run over the allowed range (to be specified shortly) of the quantum number $m$. This equation agrees with the decomposition (\ref{decomposition appendix initial  TT+long}), (\ref{decomposition appendix initial long only}) with the following identifications:
\begin{align}
    \Psi^{(\text{TT})}_{\mu \nu} &=\sum_{n=2}^{\infty}\sum_{\sigma = \pm }a^{\sigma}_n \, \psi^{\sigma}_{n|\mu\nu}~, \\
    \Psi^{(\text{long} | \text{vec-spinor})}_{\mu \nu} &=\sum_{m}\sum_{\sigma = \pm }\sum_{j=1,2}b^{(j)\sigma}_m \, v^{(j)\sigma}_{m|\mu\nu}~, \label{equation to prove -- long vec-spinor}\\
    \Psi^{(\text{long} | \text{spinor})}_{\mu \nu} &=\sum_{m}\sum_{\sigma = \pm }\sum_{j=1,2}c^{(j)\sigma}_m \, \lambda^{(j)\sigma}_{m|\mu\nu}~. \label{equation to prove -- long spinor}
\end{align}
The first equation above is simply the statement that a TT tensor-spinor on $S^4$ can be expressed as a mode sum in terms of TT tensor-spinor spherical harmonics. The main purpose of this Appendix is to verify that the decomposition (\ref{decomposition appendix initial  TT+long})-(\ref{decomposition appendix longitudinal spinor}) is the most general one, and in order to do so we need to show that (\ref{equation to prove -- long vec-spinor}), (\ref{equation to prove -- long spinor}) are respectively equivalent to (\ref{decomposition appendix longitudinal vec-spinor}), (\ref{decomposition appendix longitudinal spinor}). The longitudinal case is more subtle than the TT case because certain vector-spinor and spinor modes are excluded from (\ref{decomposition appendix longitudinal vec-spinor}) and (\ref{decomposition appendix longitudinal spinor}), and we will shortly explain why this is the case. Moreover, we will explicitly write down the longitudinal harmonics $v^{(j)\sigma}_{m|\mu\nu}$ and $\lambda^{(j)\sigma}_{m|\mu\nu}$, and we will verify that $ \Psi^{(\text{long} | \text{vec-spinor})}_{\mu \nu}$  and $ \Psi^{(\text{long} | \text{spinor})}_{\mu \nu}$ can be expressed as in (\ref{decomposition appendix longitudinal vec-spinor}) and (\ref{decomposition appendix longitudinal spinor}), respectively. 


\subsection{Constructing longitudinal spherical harmonics }

\paragraph{Longitudinal harmonics built out of spinors.} Consider the spinor spherical harmonics on $S^4$, $\psi_{m}^{\pm}$ \cite{Camporesi:1995fb}, where the quantum number $m$ takes values in $ \{0,1,\dots \}$ (let $s= \tfrac{1}{2}$ in (\ref{dirac spin s spectrum})). Using these spinors, we can construct the longitudinal symmetric tensor-spinor spherical harmonics $\lambda^{(j)\pm}_{m|\mu\nu}$, with $j=1,2\,$, which satisfy
\begin{align}
    \slashed{\nabla}\lambda^{(j)\pm}_{m|\mu\nu} = i \, \xi^{(j)\pm}_m \, \lambda^{(j)\pm}_{m|\mu\nu} ~,
\end{align}
and the eigenvalues are
\begin{align}
    \xi^{(j)\pm}_m = i\,(-1)^{j}\sqrt{8-(m+2)^2} ~.
\end{align}
They can be explicitly written down as
\begin{align}
     \lambda^{(j)\pm}_{m|\mu\nu} = \Big( \nabla_{(\mu} \nabla_{\nu)} + \alpha^{(j)\pm}_m \gamma_{(\mu} \nabla_{\nu)} + \beta^{(j) \pm}_m  g_{\mu \nu} \Big) \psi_{m}^{\pm} ~,
\end{align}
where
\begin{align}
    &\alpha^{(j) \pm}_m = \frac{1}{2} \Big(  \mp i (m+2) + i \xi^{(j)\pm}_m \Big) ~, \\
    &\beta^{(j) \pm}_m=\frac{1}{4}   \frac{\pm (m+2)+ 3 \xi^{(j)\pm}_m}{\pm(m+2)-\xi^{(j)\pm}_m} ~. 
\end{align}
It is easy to see that the longitudinal modes with $m=0$ vanish identically. Indeed, given that $\psi^{\pm}_0$ are Killing spinors (i.e. they satisfy $\nabla_\mu \psi^{\pm}_0 = \pm \frac{i}{2}\gamma_\mu \psi^{\pm}_0$), one finds
\begin{align}
     \lambda^{(j)\pm}_{m=0|\mu\nu}= 0 ~.
\end{align}
Thus, the modes $\lambda^{(j)\pm}_{m|\mu\nu}$ exist only for $m \geq 1$. Using our explicit expression for the longitudinal harmonics, we can rewrite (\ref{equation to prove -- long spinor}) as 
\begin{align}
    \Psi^{(\text{long}|\text{spinor})}_{\mu \nu} = &~ \nabla_{(\mu} \nabla_{\nu)}\sum_{m = 1}^\infty \sum_{\sigma= \pm} \sum_{j=1,2}c^{(j)\sigma}_m \, \psi_{m}^{\sigma} \nonumber + \gamma_{(\mu} \nabla_{\nu)}\sum_{m=1}^\infty \sum_{\sigma= \pm} \sum_{j=1,2}c^{(j)\sigma}_m \, \alpha^{(j)\sigma}_m \, \psi_{m}^{\sigma} \nonumber \\
    & + g_{\mu \nu} \sum_{m=1}^\infty \sum_{\sigma= \pm} \sum_{j=1,2}c^{(j)\sigma}_m \, \beta^{(j)\sigma}_m \, \psi_{m}^{\sigma} ~.  \label{eq appendix getting there}
\end{align}
From this equation, it is clear that we expressed $\Psi^{(\text{long}|\text{spinor})}_{\mu \nu} $ in the desired form (\ref{decomposition appendix longitudinal spinor}), and at the same time we verified that all spinors that appear in (\ref{decomposition appendix longitudinal spinor}) are orthogonal to Killing spinors $\psi^{\pm}_{0}$.
\newline\newline
To complete the analysis of $\Psi^{(\text{long}|\text{spinor})}_{\mu \nu}$, we have to show that the spinor $\Lambda$ in (\ref{decomposition appendix longitudinal spinor}) is orthogonal to all spinor potentials. In other words, we must show that $\Lambda$ is orthogonal to all spinor spherical harmonics $\psi^{\pm}_{m=1}$ (see Appendix \ref{App: conformal Killing vec-spin}). A straightforward calculation shows that certain modes vanish identically, namely
\begin{equation}
    \lambda^{(1)+}_{m=1|\mu\nu} = 0 = \lambda^{(2)-}_{m=1|\mu\nu} ~,
\end{equation}
where we used the fact that spinor potentials $\psi^\pm_{m=1} \propto \phi^{\pm}$ satisfy (\ref{def: spinor potentials very nice}). We further observe that $\beta^{(j=1)-}_{m=1} = 0 = \beta^{(j=2)+}_{m=1}$, and consequently
\begin{equation}
    \lambda^{(j=1)-}_{m=1|\mu\nu}= \Big( \nabla_{(\mu} \nabla_{\nu)} + \alpha^{(j=1)-}_{m=1} \gamma_{(\mu} \nabla_{\nu)}   \Big) \psi_{m=1}^{-}~,
\end{equation}
\begin{equation}
     \lambda^{(j=2)+}_{m=1|\mu\nu}= \Big( \nabla_{(\mu} \nabla_{\nu)} + \alpha^{(j=2)+}_{m=1} \gamma_{(\mu} \nabla_{\nu)}   \Big) \psi_{m=1}^{+}~,
\end{equation}
i.e. these modes do not contain terms $\propto g_{\mu \nu}$. From these observations, we conclude that the terms proportional to $g_{\mu \nu}$ in (\ref{eq appendix getting there}) do not include any spinor potentials, as
\begin{align}
    \Psi^{(\text{long}|\text{spinor})}_{\mu \nu} =&~ \dots
    + g_{\mu \nu} \sum_{m=2}^\infty \sum_{\sigma= \pm} \sum_{j=1,2}c^{(j)\sigma}_m \, \beta^{(j)\sigma}_m \, \psi_{m}^{\sigma} ~,
\end{align}
where $\dots$ contains the terms $\propto \nabla_{(\mu} \gamma_{\nu)}$ and $\propto \gamma_{(\mu} \nabla_{\nu)}$. Comparing with (\ref{decomposition appendix longitudinal spinor}), we conclude that the spinor $\Lambda$ is orthogonal to all spinor potentials.

\paragraph{Longitudinal harmonics built out of TT vector-spinors.} Consider the TT vector-spinor spherical harmonics on $S^4$, $\psi_{m|\mu}^{\pm}$ \cite{Letsios:2022tsq}, where the quantum number $m$ takes values in $ \{1,2,\dots \}$ (let $s= \tfrac{3}{2}$ in (\ref{dirac spin s spectrum})). Using these TT vector-spinors, we can construct the longitudinal symmetric tensor-spinor spherical harmonics $v^{(j)\pm}_{m|\mu\nu}$, with $j=1,2\,$, which satisfy
\begin{equation}
    \slashed{\nabla}v^{(j)\pm}_{m|\mu\nu} = i \, l^{(j)\pm}_m \, v^{(j)\pm}_{m|\mu\nu} ~,
\end{equation}
and the eigenvalues are
\begin{equation}
     l^{(j)\pm}_m = (-1)^{j}\sqrt{(m+2)^2-5} ~.
\end{equation}
They can be explicitly written down as
\begin{align}
     v^{(j)\pm}_{m|\mu\nu}=  \nabla_{(\mu} \psi_{m|\nu)}^{\pm} + A^{(j)\pm}_m \gamma_{(\mu} \psi_{m|\nu)}^{\pm} ~,
\end{align}
where
\begin{align}
    A^{(j)\pm}_m = \frac{1}{2}  \left(i\,l^{(j)\pm}_m  \mp i (m+2)  \right)~.
\end{align}
Similarly to the previous case of $\Psi^{(\text{long}|\text{spinor})}_{\mu \nu}$, using our explicit expressions for the longitudinal modes $ v^{(j)\pm}_{m|\mu\nu}$, we can show that $\Psi^{(\text{long}|\text{vector-spinor})}_{\mu \nu}$ (\ref{equation to prove -- long vec-spinor}) can be re-expressed in the desired form (\ref{decomposition appendix longitudinal vec-spinor}). Moreover, it follows from our analysis that the vector-spinor  $\lambda^{(\text{TT})}_{\nu}$ in (\ref{decomposition appendix longitudinal vec-spinor}) is orthogonal to all TT vector-spinor modes 
$\psi_{m=1|\nu}^{\pm}$, which are Killing vector-spinors (\ref{def: Killing vec-spinors very nice}) on $S^4$.

\bibliographystyle{JHEP}
\bibliography{bib}

\end{document}